\documentclass[10pt,a4paper,twoside]{article}
\usepackage{epsfig}
\usepackage{baltlat6}
\usepackage{array}
\usepackage{longtable}
\usepackage{wrapfig}
\begin{document}
\ \ \vspace{0.5mm} \setcounter{page}{273} \vspace{8mm}

\titlehead{Baltic Astronomy, vol.\,22, 273--296, 2013}

\titleb{A NEW EMPIRICAL METALLICITY CALIBRATION FOR\\ VILNIUS PHOTOMETRY }

\begin{authorl}
\authorb{S. Barta\v{s}i\={u}t\.e}{},
\authorb{S. Raudeli\={u}nas}{}
and
\authorb{J. Sperauskas}{}
\end{authorl}

\moveright-3.2mm
\vbox{
\begin{addressl}
\addressb{}{Astronomical Observatory of Vilnius University,
\v{C}iurlionio 29, Vilnius,\\ LT-03100, Lithuania}
\end{addressl} }

\submitb{Received: 2013 August 7; accepted 2013 November 29}

\begin{summary} We present a new calibration of the seven-color
Vilnius system in terms of [Fe/H], applicable to F--M stars in the
metallicity range $-2.8\leq$[Fe/H]$\leq+0.5$. We employ a purely
empirical approach, based on $\sim$\,1000 calibrating stars with
high-resolution spectroscopic abundance determinations. It is shown
that the color index $P$--$Y$ is the best choice for a most accurate
and sensitive abundance indicator for both dwarf and giant stars.
Using it, [Fe/H] values can be determined with an accuracy of
$\pm$\,0.12 dex for stars of solar and mildly sub-solar metallicity
and $\pm0.17$ dex for stars with [Fe/H]\,$<$\,--1. The new
calibration is a significant improvement over the previous one used
to date.
\end{summary}

\begin{keywords} techniques: photometric -- stars: fundamental
parameters -- stars: abundances -- stars: metallicity calibration

\end{keywords}

\resthead{Photometric metallicity calibration}{ S. Barta\v{s}i\={u}t\.e,
S. Raudeli\={u}nas, J. Sperauskas}

\sectionb{1}{INTRODUCTION}

Metallicity estimation from broad- and intermediate-band photometry
is a very practical and efficient means to obtain metal abundances
for large samples of faint stars. Such methods are based on the
sensitivity of stellar color indices to photospheric abundances over
a relatively wide wavelength range. Techniques for deriving
metallicity parameters [Fe/H] were developed and are used in a
number of photometric systems, e.g., {\it Sloan}, {\it uvby}, {\it UBV},
{\it DDO}, etc.

The metal abundance sensitivity of the seven-color {\it Vilnius}
system {\it UPXYZVS} was first demonstrated in the papers by
Bartkevi{\v c}ius \& Strai\v{z}ys (1970a,b). This system (for its
details, see Strai\v{z}ys 1992) includes several color indices in
the region $\sim3200-4800$~\AA, shown to be most effective at
differentiating the metallicity effect: $U$--$X$, $U$--$Y$,
$P$--$X$, $P$--$Y$ and $X$--$Y$. Later on, Bartkevi{\v c}ius \&
Sperauskas (1983; hereafter BS83) calibrated empirically three of
the above-quoted color indices in terms of [Fe/H] by using 133 field
stars with known values of metallicity, but only half of these stars
then had [Fe/H] values from high-dispersion spectroscopy. Their
calibration of $P$--$Y$ and $X$--$Y$ applies to dwarf stars of
spectral types F0--K4 in the metallicity range from $+0.5$ to $-2.5$
dex and that of $P$--$X$, $P$--$Y$ and $X$--$Y$ applies to giant
stars of types F8--M4 in the [Fe/H] range from $+0.5$ to $-3.0$ dex.
The order of accuracy of the BS83 calibration is about $\pm0.20$ dex
in [Fe/H].

Since the publication of BS83, a fair amount of observational data
have been accumulated, both in the {\it Vilnius} system and from
high-dispersion spectroscopy. This permits a much-improved
recalibration of {\it Vilnius} photometry. Such a need has become
increasingly important in the light of recent photometric CCD
surveys in the {\it Vilnius} system, which have substantially
increased the quality and volume of photometric data available for
the Milky Way clusters and field stars (e.g., Barta{\v s}i{\= u}t{\.
e} et al. 2011; Zdanavi{\v c}ius et al. 2011, 2012; Strai{\v z}ys et
al. 2013).

We have therefore undertaken a reevaluation of the original BS83
calibration and provide in this paper its updated version which
keeps the same empirical approach as that used in BS83 but expands a
calibrating sample to include nearly 1000 F--M stars (a seven-fold
increase in sample size) with more recent [Fe/H] values from
high-dispersion spectra and thus to ensure a better accuracy of
metallicity estimation.

The layout of the paper is as follows. In Section~2 we describe the
stellar sample and the data used for metallicity calibration.
Section~3 describes details of the empirical method used and
presents a new calibration for dwarfs and giant stars. In this
section we compare the accuracy of the updated and original versions
of calibration and discuss their implications. Conclusions and
recommendations are summarized in Section~4.

\vskip2mm

\sectionb{2}{THE SAMPLE}

In any attempted metallicity calibration, it is critically important
to have an extensive and homogeneous sample of spectroscopically
determined abundances for stars for which there are also accurate
{\it Vilnius} photometric data.

As a source catalog for {\it Vilnius} photometry the latest updated
version of the {\it General Photometric Catalogue of Stars Observed
in the Vilnius System} (Kazlauskas 2010) was used, which contains
compilations from the published sources for $\sim$10\,000 stars and
supersedes the previous catalog by Strai\v{z}ys \& Kazlauskas
(1993). Only stars which have both the {\it Hipparcos} parallaxes
and the metallicity determinations from high-dispersion spectra were
extracted from this catalog. Where more than one source was
available for a given star, the results of photometry were averaged.
If the differences between the values of color indices from
different sources exceeded 0.02 mag (in the case of non-variable
stars), such averaged color indices were treated as inaccurate and
not used in metallicity calibration.

The spectroscopic metallicity determinations were taken from the
PASTEL catalog by Soubiran et al. (2010) and its updated
version\footnote{~http://vizier.u-strasbg.fr/viz-bin/VizieR?-source=B/pastel}.
The catalog supersedes the two previous versions (Cayrel de Strobel
et al. 1997, 2001) and provides the most recent compilation of
[Fe/H] determinations obtained from detailed analysis of high
resolution, high $S/N$ spectra, together with the atmospheric
parameters $T_{\rm e}$ and $\log g$, spectral types and
bibliographic references. For a small number of stars, [Fe/H]
determinations from high-dispersion spectra were found in the very
recent literature sources not yet included in the updated PASTEL
version (e.g., Afsar et al. 2012). Since the metallicity
determinations in the PASTEL compilations come from a variety of
sources it was not possible to claim zero point offsets between the
metallicity scales. Therefore, we took from PASTEL the values of
[Fe/H] without any change in zero-points. In the case of multiple
[Fe/H] determinations, we adopted [Fe/H] values averaged over
different sources (no weights were attached to authors of [Fe/H]
estimates, but only sources after 1990 were used in averaging). In
the case of considerable discrepancy in the published values of
[Fe/H], mainly because of different effective temperatures assumed,
we avoided the inclusion of such stars. We note that metallicity
determinations by McWilliam (1990), given for a large number of G--K
giant stars in our sample (272), were found to be systematically by
0.11 dex lower, on average, than those of the many other sources
(see, also, a remark on this point by Liu et al. 2007). Since the
McWilliam stars cover mostly the range [Fe/H]\,$>$\,$-0.5$, in which
the majority of our sample stars had [Fe/H] values from other
sources, we did not attempt to tie the abundance scale of this
particular source to some known zero-point, but excluded this source
from averaging and from further calibration procedures.

\begin{figure}[!t]
\centerline{\psfig{figure=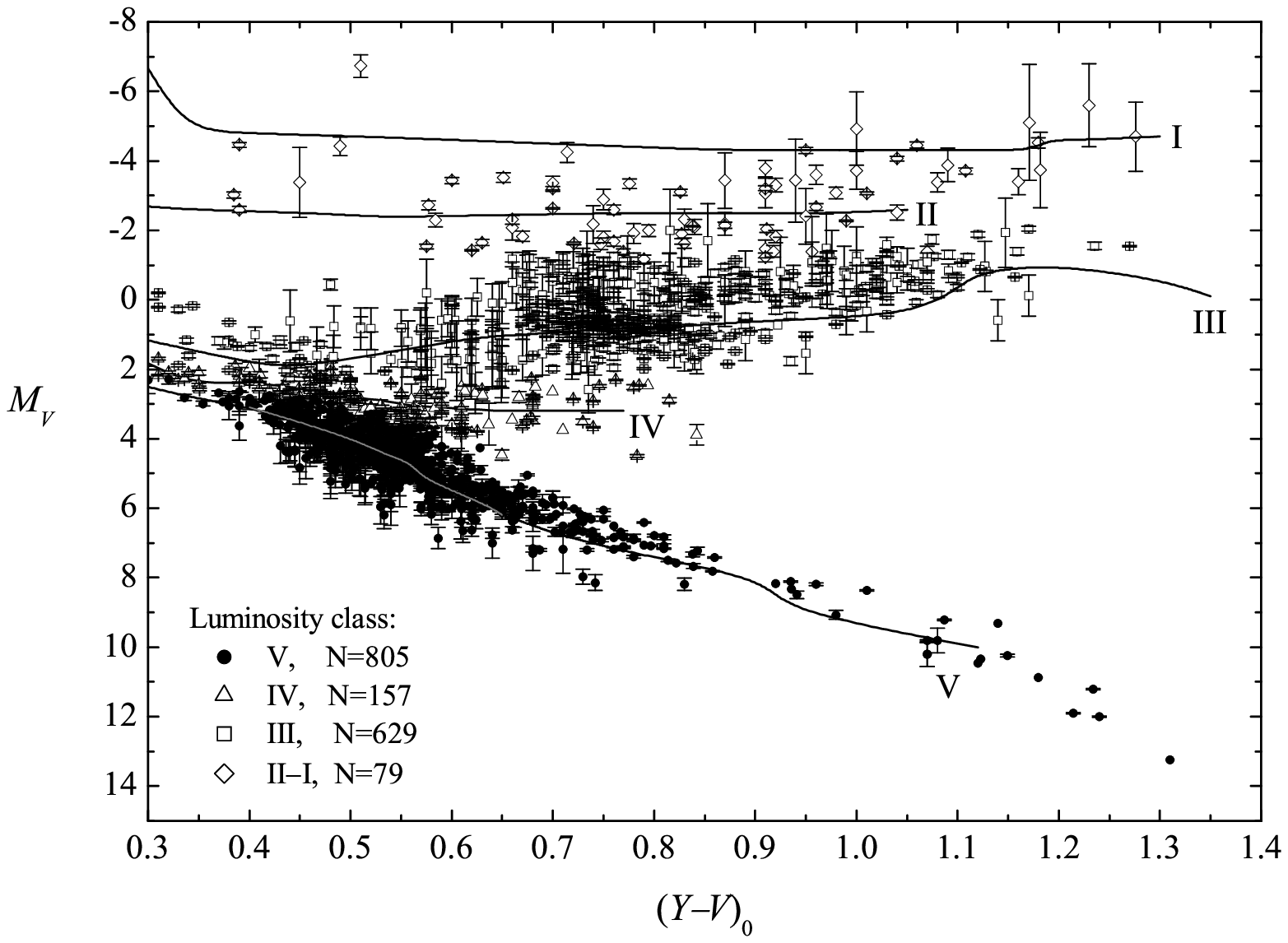,width=106mm,angle=0,clip=}}
\vskip0.5mm \captionb{1}{$M_V$ vs. $(Y$--$V)_0$ diagram for 1666
stars with metallicity estimates from high-dispersion spectra.
Absolute magnitudes are calculated from Hipparcos parallaxes, with
the exception of 66 stars with unacceptably large parallax errors,
mainly of luminosity class III to I, for which $M_V$ were derived
from {\em Vilnius} photometry. The mean relations for solar
composition stars of luminosity classes V to I, taken from
Strai\v{z}ys (1992), are shown as continuous lines.}
\end{figure}

The use of the Hipparcos catalog (new reduction, van Leeuwen 2007)
allowed us to reliably determine distances to the calibrating stars
and to check an assessment of their luminosity classes. The
knowledge of accurate distances were critically important to see
whether there was the need for correction of colors due to
interstellar reddening or not. The stars lying at distances less
than 40 pc were assumed to be free of reddening. For stars at
greater distances, the values of interstellar reddening were
determined using regular techniques of photometric classification in
the {\it Vilnius} system (see, e.g., Bartkevi{\v c}ius \&
Lazauskait{\. e} 1996 for the principle of this technique). In the
case of nonzero reddening, the colors of stars were dereddened using
the color excess ratios taken from Kurilien{\. e} \& S{\= u}d{\v
z}ius (1974) and Bartkevi{\v c}ius \& Sviderskien{\. e} (1981) for
Population I and II stars, respectively.

A total of 1666 stars were found which have both {\it Vilnius}
photometry and metallicity estimates from high-dispersion spectra.
The $M_V$ vs. $(Y$--$V)_0$ diagram for these stars is displayed in
Figure 1. The stars of luminosity class IV (157 stars) and
luminosity class II--I (79 stars) are not the subject of current
metallicity calibrations and have therefore been omitted from our
subsequent analysis, leaving a sample of 805 dwarf stars and 625
giant stars, which cover a range of $Y$--$V$ color from 0.35 to 1.3,
or, spectral classes from F0 to M4. The distributions of these stars
by metallicity and the values of interstellar reddening $E_{Y-V}$
are shown in Figure~2 (unshaded histograms). As can be seen from the
histograms, the range of metallicities covered is
$-2.8$\,$\leq$\,[Fe/H]\,$\leq$\,$+0.4$. The overwhelming majority of
the dwarfs are nearby and unreddened, whereas a significant fraction
of the giants are slightly reddened, having color excesses $E_{Y-V}$
mostly less than 0.05 mag. Very few stars have slightly larger
reddening values, but not exceeding 0.10 mag.

\begin{figure}[!t]
\begin{center}
 \resizebox{6.1cm}{!}{\includegraphics{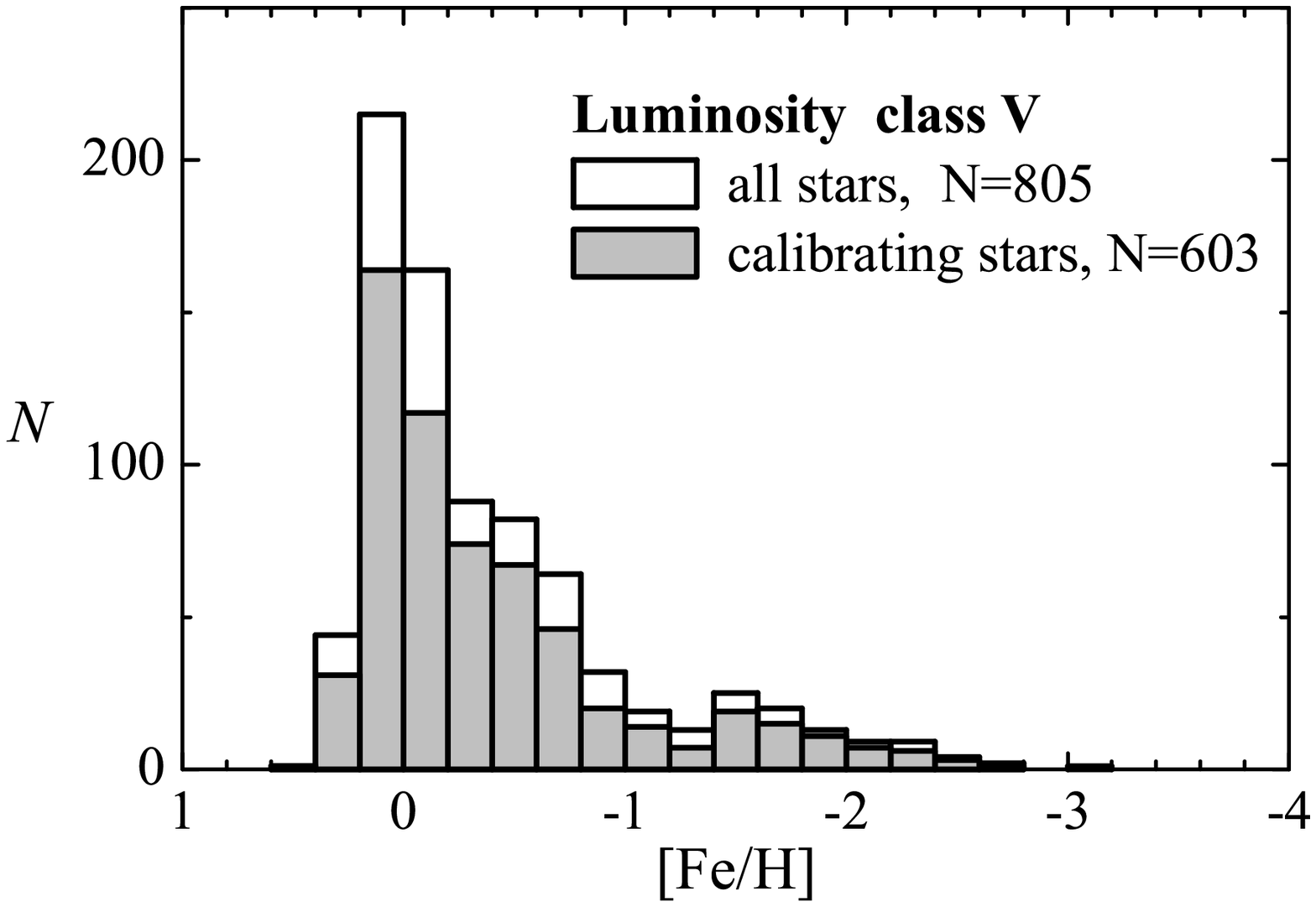}}
 \resizebox{6.1cm}{!}{\includegraphics{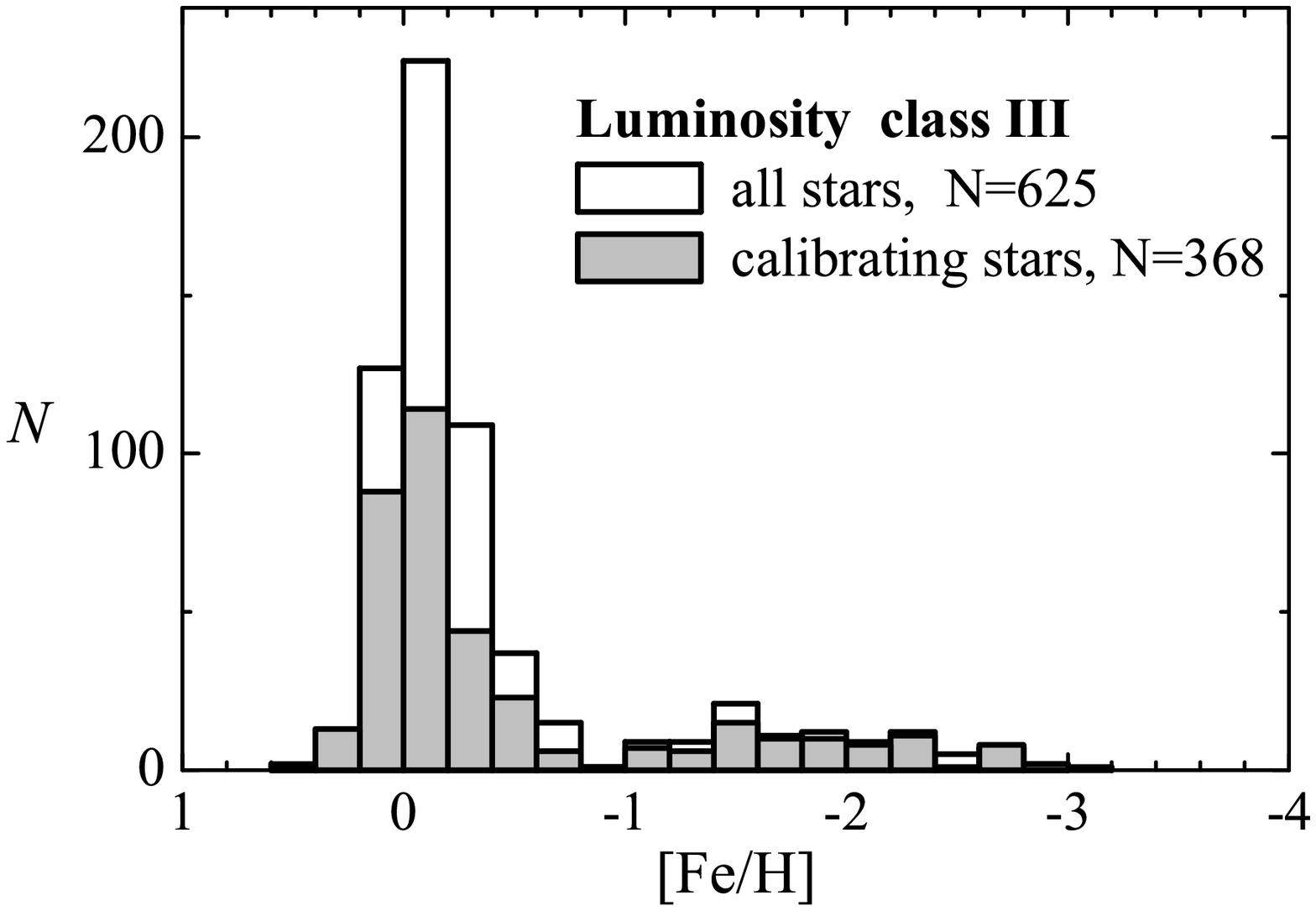}}
\vskip2mm
 \resizebox{6.1cm}{!}{\includegraphics{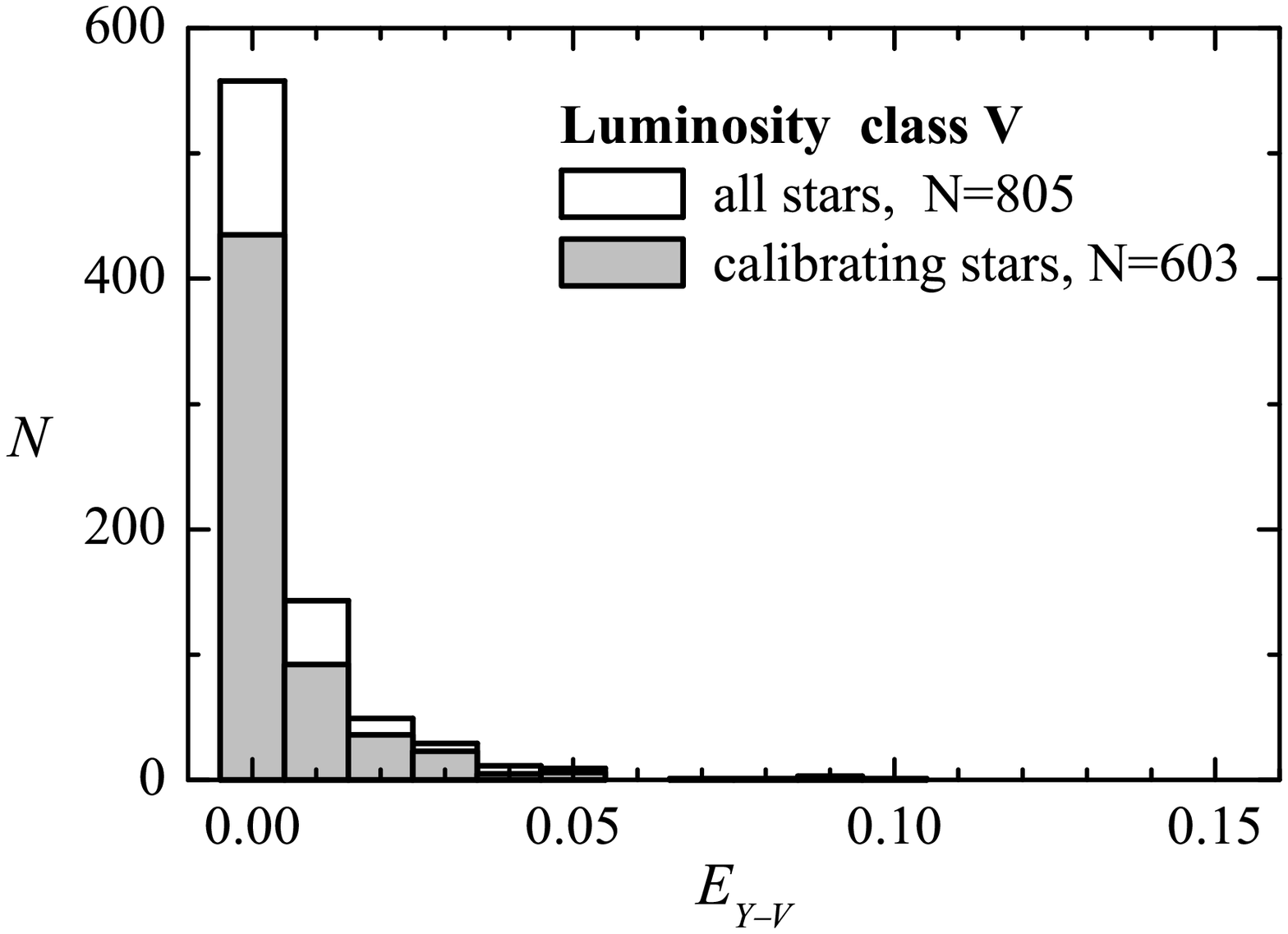}}
 \resizebox{6.1cm}{!}{\includegraphics{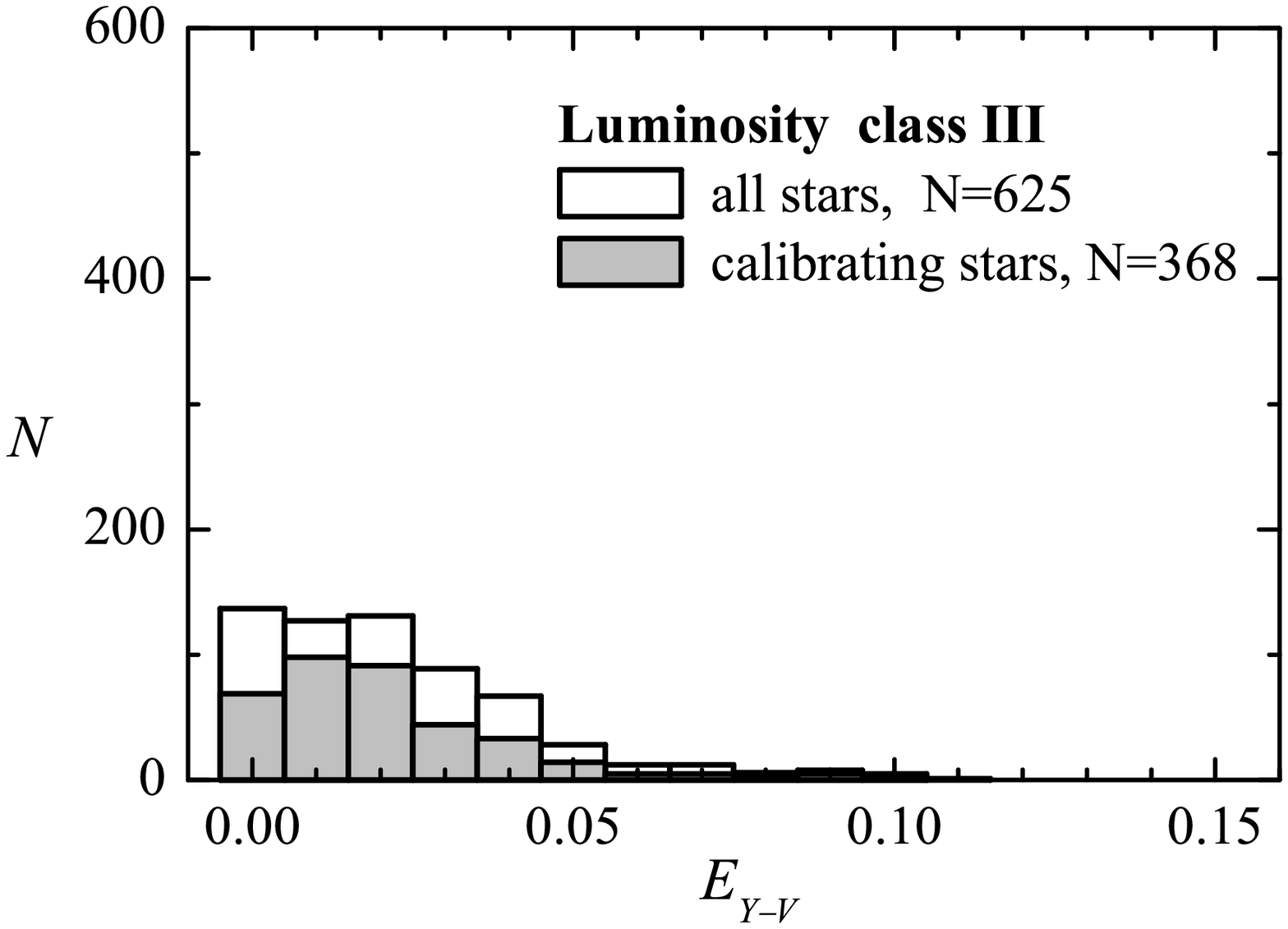}}
 \end{center}
 \vspace*{-0.2cm}
 \captionb{2}{Distributions of dwarf and giant stars by metallicity (top panels)
 and by values of interstellar reddening, $E_{Y-V}$ (bottom panels).
 Unshaded histograms represent all stars in the sample, whereas shaded histograms
 indicate calibrating stars left after removal of binaries,
 variables and stars with inaccurate photometry or with discrepant values of spectroscopic [Fe/H].}
\end{figure}

Particular attention has been paid to removing unresolved binaries
from a sample of calibrating stars, which primarily affect
photometric measurements, making the binary system appear brighter
and redder. Consequently, certain combinations of primary and
secondary stars can give quite highly erroneous photometric
estimates of metallicities. We have searched for known spectroscopic
binaries by making a prior cross-checking of our sample of dwarf and
giant stars with the catalog of spectroscopic binaries by Pourbaix
et al. (2004--2009) and with the catalogs of radial velocities by
Nidever et al. (2002), Famaey et al. (2005) and Nordstroem et al.
(2004). Also, the Simbad database and the {\it Hipparcos} catalog
entries for all of the stars were checked to avoid the inclusion of
any double stars. The known binaries and radial-velocity variables
comprise 20\% and 17\% of our subsamples of dwarf and giant stars,
respectively. These were excluded from calibration procedures.
However, some unrecognized double stars can almost certainly remain
in our cleaned sample.

The known variable stars in the sample have been searched by
cross-checking with the {\it General Catalog of Variable Stars}
(Samus et al. 2007--2012); these constitute 17\% of the sample, with
nearly one third of them known also as spectroscopic binaries. Only
those which were known to exhibit low-level ($\leq 0.02$ mag)
variability were admitted to the final sample to be used in
metallicity calibration.

After excluding known binaries, variable stars, carbon and Ba stars
(most of which are also spectroscopic binaries) and also stars with
less accurate photometry (differences in color indices between
different sources $\geq$\,0.03 mag) or discrepant values of [Fe/H],
we retained in our final sample for metallicity calibration 603
dwarfs and 368 giants. Their distributions by [Fe/H] and reddening
values $E_{Y-V}$ are shown by shaded histograms in Figure 2.

\vskip2mm

\sectionb{3}{METALLICITY CALIBRATIONS}

\subsectionb{3.1}{Method}

In this paper, we used the same empirical approach as that given in
BS83, which is based on the method put forward by Bond (1980) for
metallicity calibration of the {\it Str{\"o}mgren} $m_1$ index.
Here, we will only briefly describe the main points of the method.

Line weakening or strengthening due to the effects of metallic lines
in stellar atmospheres (line blanketing) can be measured by the
color excess $\delta (CI)$ defined as the difference between a
star's metallicity-sensitive color index $CI$ and the same color
index that this star would have if its metallicity were exactly
solar ([Fe/H]=0.00); we shall label the latter $(CI)_{\rm n}$. In a
two color diagram with a temperature-dependent color index plotted
on the $x$-axis, $\delta (CI)$ is a height of a star's point above
(or below) the solar-composition relation. However, we cannot use a
simple linear relation between $\delta (CI)$ and [Fe/H], since the
abundance sensitivity of a color index is not the same for different
intervals of effective temperature. This effect is mostly
attributable to a progressive increase of line blanketing with
decreasing temperature. Instead, a more relevant quantity
$\delta^\prime (CI)$, introduced originally by Bond (1980), can be
used, which was also employed in the calibration by BS83. In a two
color diagram, $\delta^\prime (CI)$ defines the above described
excess $\delta (CI)$ in units of the distance $\delta (CI)_{\rm
max}$ between the solar composition relation and the line of some
fixed maximal metal-deficiency $(CI)_{\rm max}$ at the same
temperature:
      \begin{equation}
        \delta^\prime (CI)={{\delta (CI)}\over{\delta (CI)_{\rm
        max}}}\,,
      \end{equation}
where $\delta(CI)=(CI)_{\rm n}-(CI)$ and $\delta(CI)_{\rm
        max}=(CI)_{\rm n}-(CI)_{\rm max}$. Having the values of
$\delta^\prime (CI)$ calculated for a sample of stars with accurate
spectroscopic metallicities the remaining step of calibrations is to
find a relation between $\delta^\prime (CI)$ and [Fe/H].

To calculate $\delta^\prime (CI)$ properly, we need to know the
two-color relations defining both the [Fe/H]=0 isoline and the
isoline of maximal metal-deficiency. The easiest way would be to use
stellar models computed for the {\it Vilnius} system. However,
despite improvements in theoretical models, it does not appear that
the transformations from theoretical to observational colors are yet
sufficiently reliable. A comparison in the two-color diagrams of the
loci of sequences of real stars with those from a widely used grid
of Kurucz (2001) model
atmospheres\footnote{~http://kurucz.harvard.edu/grids.html}, computed
for the {\it Vilnius} system,
 demonstrated that model colors do
not adequately match the observational data even in the case solar
metallicity stars (see Figures 4 and 6 and remarks in \S\,3.2 and
3.3). Therefore, we derived the metallicity dependent color shifts
$\delta^\prime (CI)$ empirically, i.e. relying solely on
observational data.

As in the BS83, we used three types of two-color diagrams,
($P$--$X$,\,$Y$--$V$), \hbox{($P$--$Y$,\,\,$Y$--$S$)} and
($X$--$Y$,\, $Y$--$V$), where $P$--$X$, $P$--$Y$ and $X$--$Y$ are
metallicity indicators and $Y$--$V$ and $Y$--$S$ are
effective-temperature indicators which are almost insensitive to
blanketing. We made no attempt to calibrate the color indices
including ultraviolet magnitude, $U$--$X$ and $U$--$Y$, because they
are most sensitive to the effects of luminosity. As color indices
$P$--$X$,\, $P$--$Y$ and $X$--$Y$ are also susceptible to
luminosity, the relations between $\delta^\prime (CI)$ and [Fe/H]
were obtained for dwarf and giant stars separately. Before computing
$\delta^\prime (CI)$, the observed color indices were corrected for
interstellar reddening, if needed. Throughout this paper, the
notations $(CI)_0$ or $CI$ always denote intrinsic color indices.

As a first step, a standard empirical relation for solar composition
stars in the two-color diagrams was defined using our sample stars
within $\pm0.10$ dex of the solar [Fe/H]. However, the small number
of stars earlier than F2 and later than M0 complicated the
determination of the relation near both ends of the spectral range
of interest. Therefore we selected from the catalog of {\it Vilnius}
photometry an additional number of stars with determined (mainly by
the BS83 technique) photometric metallicities within $\pm0.15$ dex
(a less restricted interval chosen to allow for the effect of errors
in photometric [Fe/H] determinations). Since their loci in the
two-color diagrams are quite compatible with those of a few
solar-metallicity stars, known from high-dispersion spectroscopy
(see Figures 4 and 6), we included these additional stars in
defining the ends of the relation, which refer to the spectral
intervals earlier than F2 ($Y$--$V$$<$\,$0.4$) and later than
$\sim$M0 \hbox{($Y$--$V$$>$\,$0.9$} for dwarfs and
$Y$--$V$$>$\,$1.2$ for giants).

Next, in the two-color diagrams, the mean line for a maximal
metal-deficiency, $(CI)_{\rm max}$, has to be defined, within which
our calibrations should be valid (i.e., [Fe/H] close to --3.0 dex).
Since in general we do not have sufficient observational data on
extremely metal-poor stars over the entire $Y$--$V$ color range we
are considering, it was not possible to define this line simply as
an upper envelope of the points in the two-color diagrams. To
overcome this problem, we adopted here the following approach.
Since, at a fixed metallicity, $\delta^\prime (CI)$ should
essentially be the same throughout the $Y$--$V$ color range, the
formula (1) can be used to extract the shifts $\delta (CI)_{\rm
max}$, and hence to calculate $(CI)_{\rm max}$, using a mean
two-color relation defined for stars of the same metal-deficiency if
only its corresponding quantity $\delta^\prime (CI)$ is known at any
(at least one) fixed color $Y$--$V$. Therefore, we chose fixed
values of $Y$--$V$ in the regions of the two-color diagrams, where
the calibrating stars cover the entire range of [Fe/H], and
estimated the empirical dependence of color indices on [Fe/H] as
illustrated in Figure 3. Then, the dependence of each color index
considered was extrapolated down to an arbitrary metallicity of $-3$
dex to

\begin{wrapfigure}[27]{r}[0pt]{64mm}
\psfig{figure=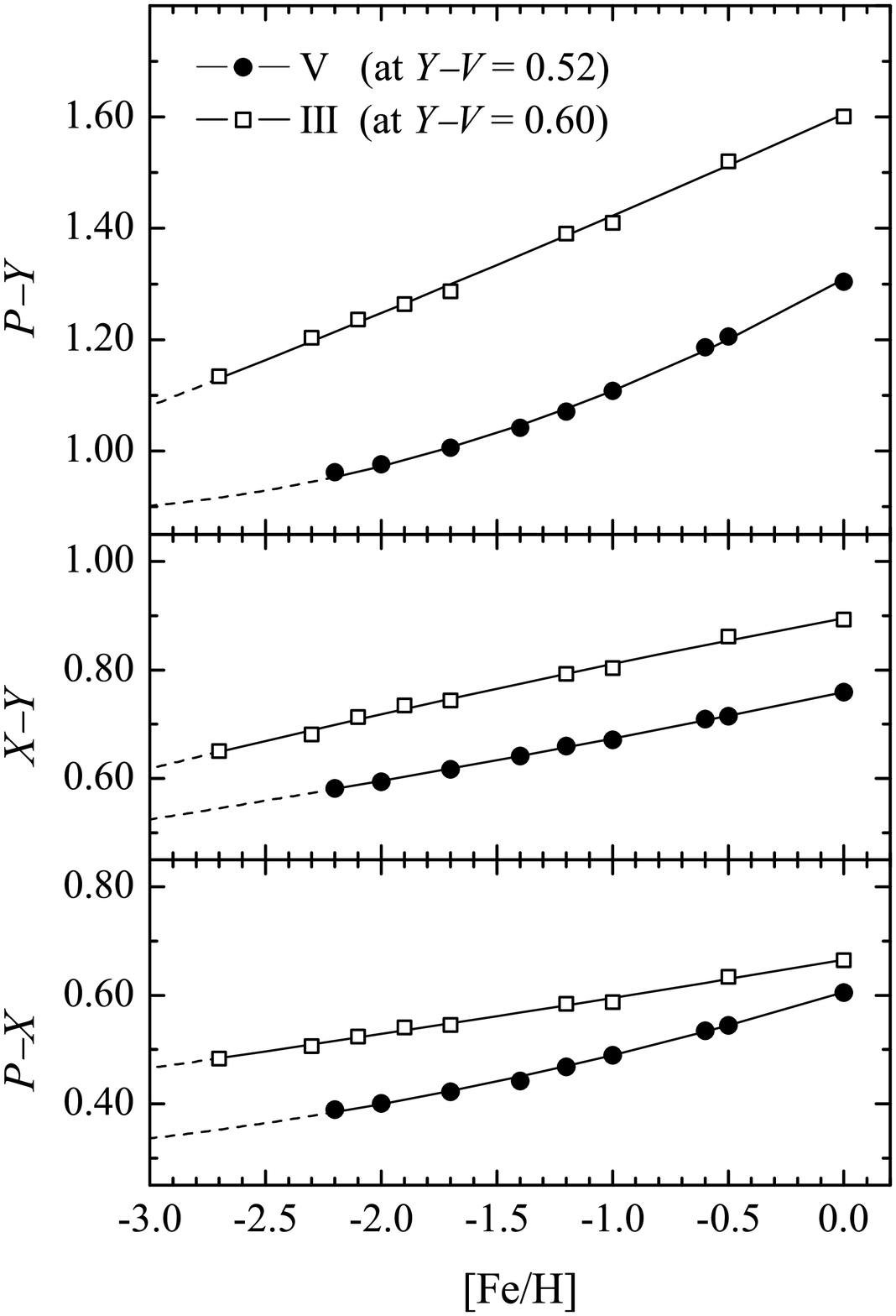,width=60mm,angle=0,clip=} \vspace{.2mm}
\captionb{3}{Dependence of the metallicity-sensitive color indices
on [Fe/H] at a fixed value of $Y$--$V$ for luminosity classes V and
III.}
\end{wrapfigure}

\noindent yield, at a given $Y$--$V$, an estimate of $\delta
(CI)_{\rm max}$ and, from formula (1), thequantities $\delta^\prime
(CI)$'s across this range of metallicity. Using the empirical lines
drawn in the two-color diagram through the points of stars with
closely similar metal-deficiency (e.g., $-0.5$, $-1.2$, $-1.5$, ...,
$-2.0$ dex), or, when the observational data were insufficient,
using only segments of such lines, and having the quantities
$\delta^\prime (CI)$ we obtained by formula (1) a set of $(CI)_{\rm
max}$ lines, which, in the ideal case, should coincide. Taking their
average position in the two-color diagram, the final $(CI)_{\rm
max}$ line was constructed over a wide range of $Y$--$V$ (or
$Y$--$S$).

The two-color relations defining the [Fe/H]=0 isoline, $(CI)_{\rm
n}$, and the isoline of maximal metal-deficiency, $(CI)_{\rm max}$,
are tabulated in Appendix. The two-color diagrams with these
empirical relations are shown  in Figures 4 and 6 for dwarfs and
giant stars, respectively, and will be briefly discussed in the next
two subsections.

\vskip2mm

\subsectionb{3.2}{Dwarf stars}

\begin{figure}[!th]
\centerline{\psfig{figure=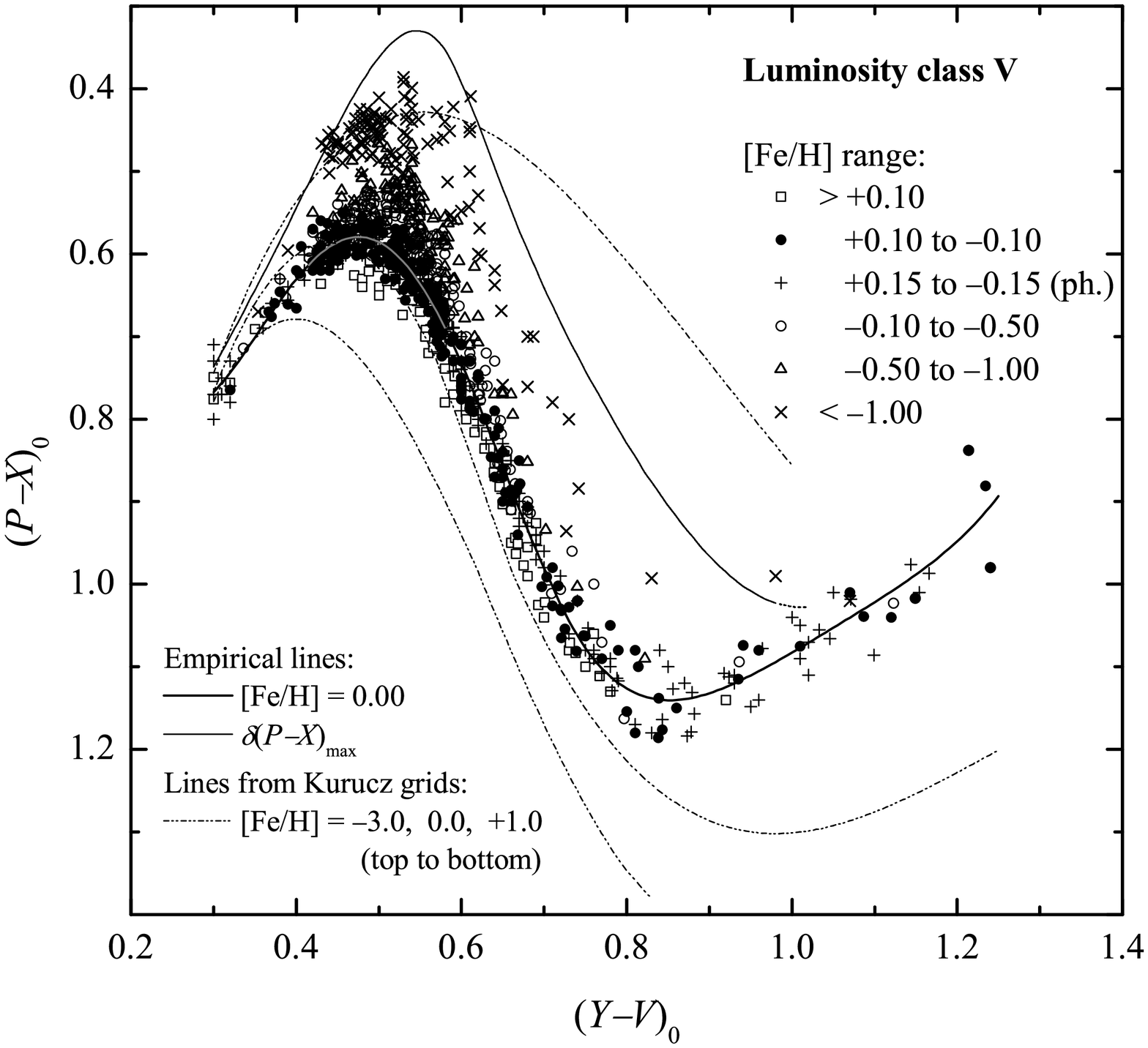,width=88mm,angle=0,clip=}}
\vskip0.5mm \captionb{4a}{$(P$--$X$,\,$Y$--$V)$ diagram for dwarf
stars. Symbols indicate stars having [Fe/H] estimates from
high-dispersion spectra, except for small thin crosses, which
indicate stars having only photometric [Fe/H]. For more
explanations, see text in \S\,3.2.}
\end{figure}
\begin{figure}[!bh]
\centerline{\psfig{figure=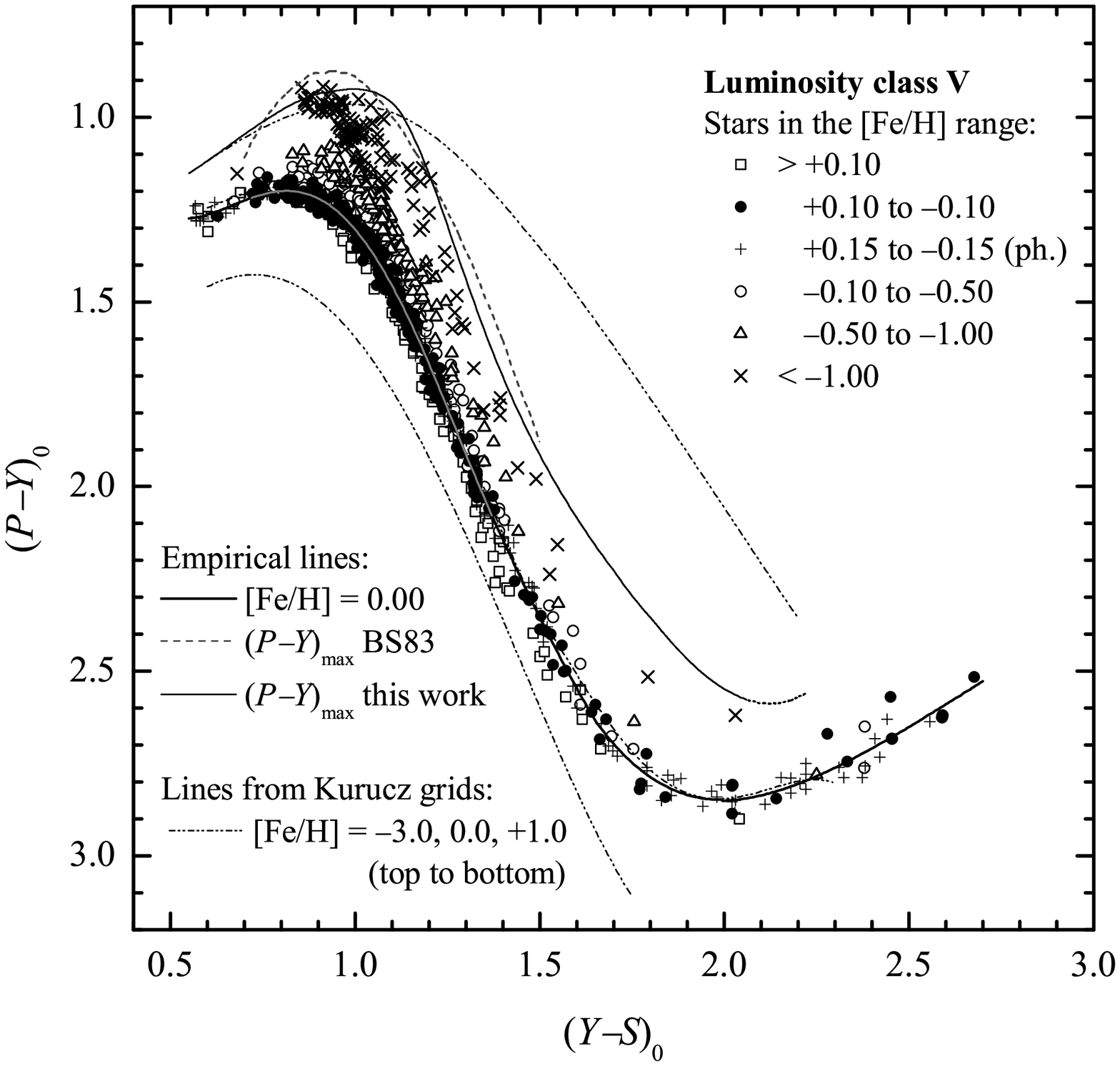,width=88mm,angle=0,clip=}}
\vskip0.5mm \captionb{4b}{$(P$--$Y$,\,$Y$--$S)$ diagram for dwarf
stars. Symbols and lines the same as in Fig.\,4\,a. Shown as dashed
line is the $(P-Y)_{\rm max}$ relation defined by BS83.}
\end{figure}

\begin{figure}[!t]
\centerline{\psfig{figure=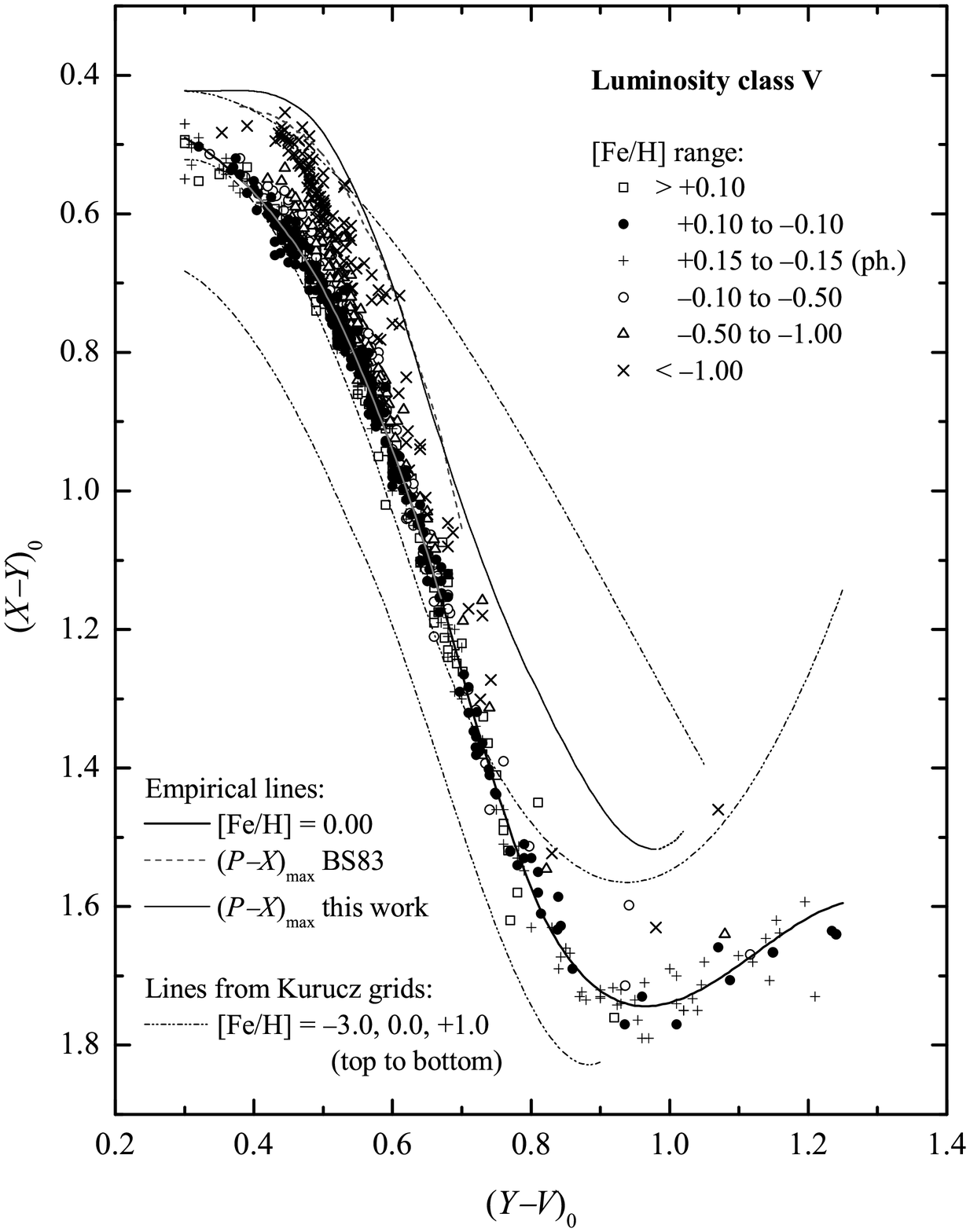,width=88mm,angle=0,clip=}}
\vskip0.5mm \captionb{4c}{$(X$--$Y$,\,$Y$--$V)$ diagram for dwarf
stars. Symbols and lines the same as in Figs.\,4\,a,b.}
\end{figure}


All the dwarfs in our sample are specifically chosen to be unevolved
stars, ensured by the requirement that their {\it Hipparcos}-based
absolute magnitudes $M_V$ should all be within the expected limits
of the main sequence (Figure 1). We divided the sample stars into
five metallicity groups and plotted the two-color diagrams in
Figures 4\,a,b,c. All symbols, except for small thin crosses,
indicate stars having [Fe/H] estimates from high-dispersion spectra.
The heavy continuous line is the mean relation for solar-metallicity
stars, drawn through the points representing stars in the range
$+0.10$\,$\geq$\,[Fe/H]\,$\geq$\,$-0.10$ (solid dots); due to
insufficient number of stars of certain spectral types, a number of
additional stars which had only photometric metallicity estimates
or, in the case when $Y$--$V$\,$>0.75$, otherwise classified as
normal chemical composition stars, were also included in the fit
(these stars are shown as small thin crosses). Thin continuous line
is the $(CI)_{\rm max}$ line defined in this paper (see \S\,3.1 for
details). Shown in Figures 4\,b,c as thin dashed line is the
$(CI)_{\rm max}$ relation used in the previous calibration by BS83.
Theoretical curves of Kurucz (2001) models (dashed-dot lines) are
shown for [Fe/H]=\,$-3.0$ (upper line), 0.0 dex (middle line) and
+1.0 dex (bottom line). We did not remove from the plots known
binaries and variable stars, therefore part of the scatter within
each group may be due to differences other than metallicity.

The rms scatter about each of the three two-color relations defined
for solar-metallicity stars of spectral types earlier than K4
($Y$--$V\leq0.75$) varies from $\pm0.02$ for $P$--$X$ and $X$--$Y$
to $\pm0.03$ for $P$--$Y$. For $Y$--$V>0.75$ (spectral types
K5--M4), the scatter is larger, $\pm$0.03 for $P$--$X$ and $\pm$0.04
for $X$--$Y$ and $P$--$Y$. We note that over most of the temperature
range considered the defined solar-metallicity relations match quite
well the mean relations given by Strai\v{z}ys (1992; Table 66) for
dwarfs of normal chemical composition (the latter relations are not
displayed in Figures 4\,a,b,c in order to prevent an overlap of the
lines). A comparison of the empirical [Fe/H]=0 relations with the
theoretical curves from Kurucz (2001) models computed for the {\it
Vilnius} system indicates a satisfactory match only in the case of
$(P$--$Y$,\,$Y$--$S)$ diagram. In the $(P$--$X$,\,$Y$--$V)$ and
$(X$--$Y$,\,$Y$--$V)$ diagrams, model colors are clearly shifted
relative to the observational sequences, with the most obvious cases
of disagreement occurring in the region of late K and M dwarfs.

\begin{figure}[!th]
\centerline{\psfig{figure=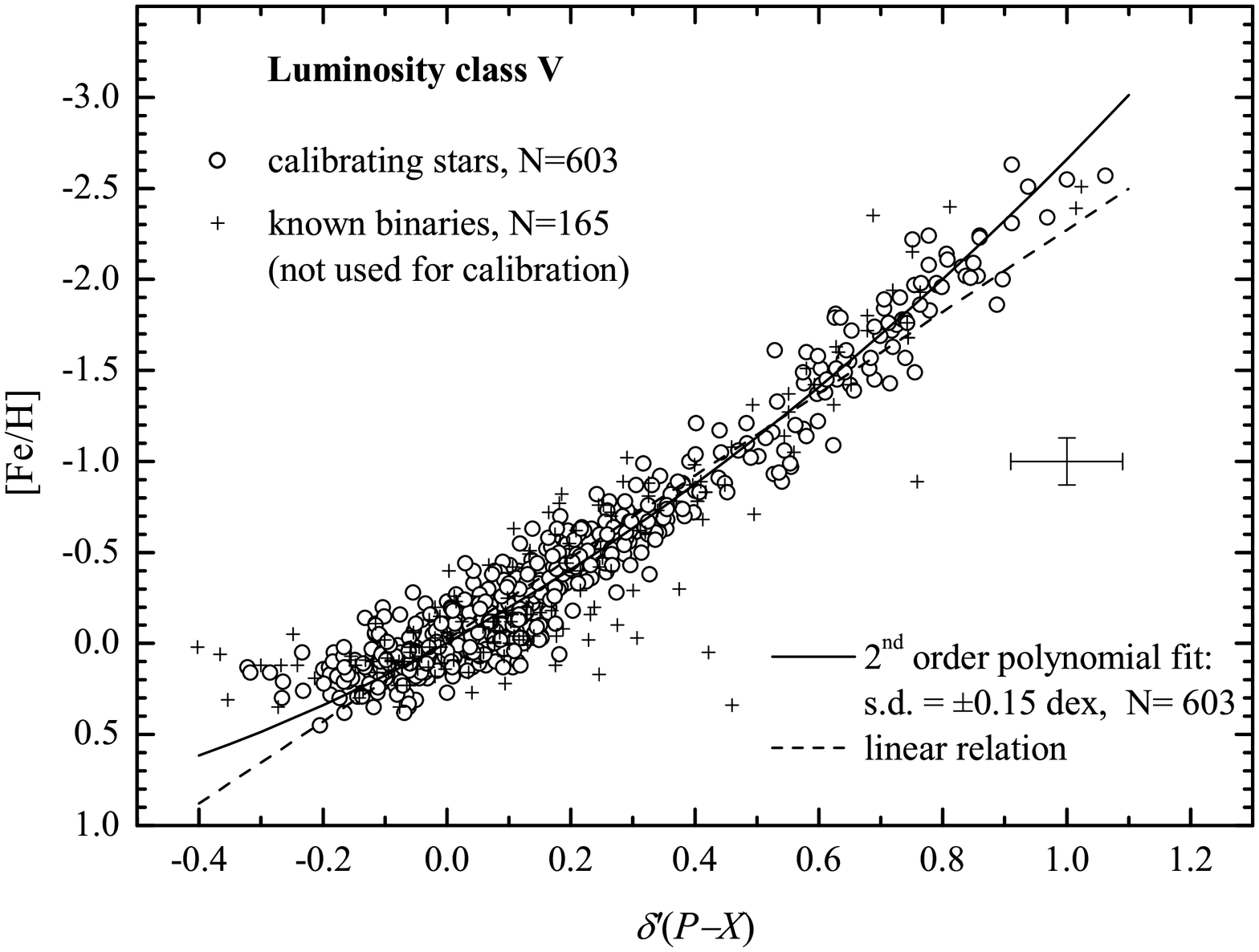,width=106mm,angle=0,clip=}}
\vskip0.5mm \captionb{5a}{$\delta^\prime(P$--$X)$ vs. [Fe/H]
relation for calibrating dwarf stars with metallicities from
high-dispersion spectroscopy. The solid line is a second-order
polynomial fit and the dashed line is a linear relation. Known
spectroscopic binaries are plotted for comparison. }
\end{figure}
\begin{figure}[!bh]
\centerline{\psfig{figure=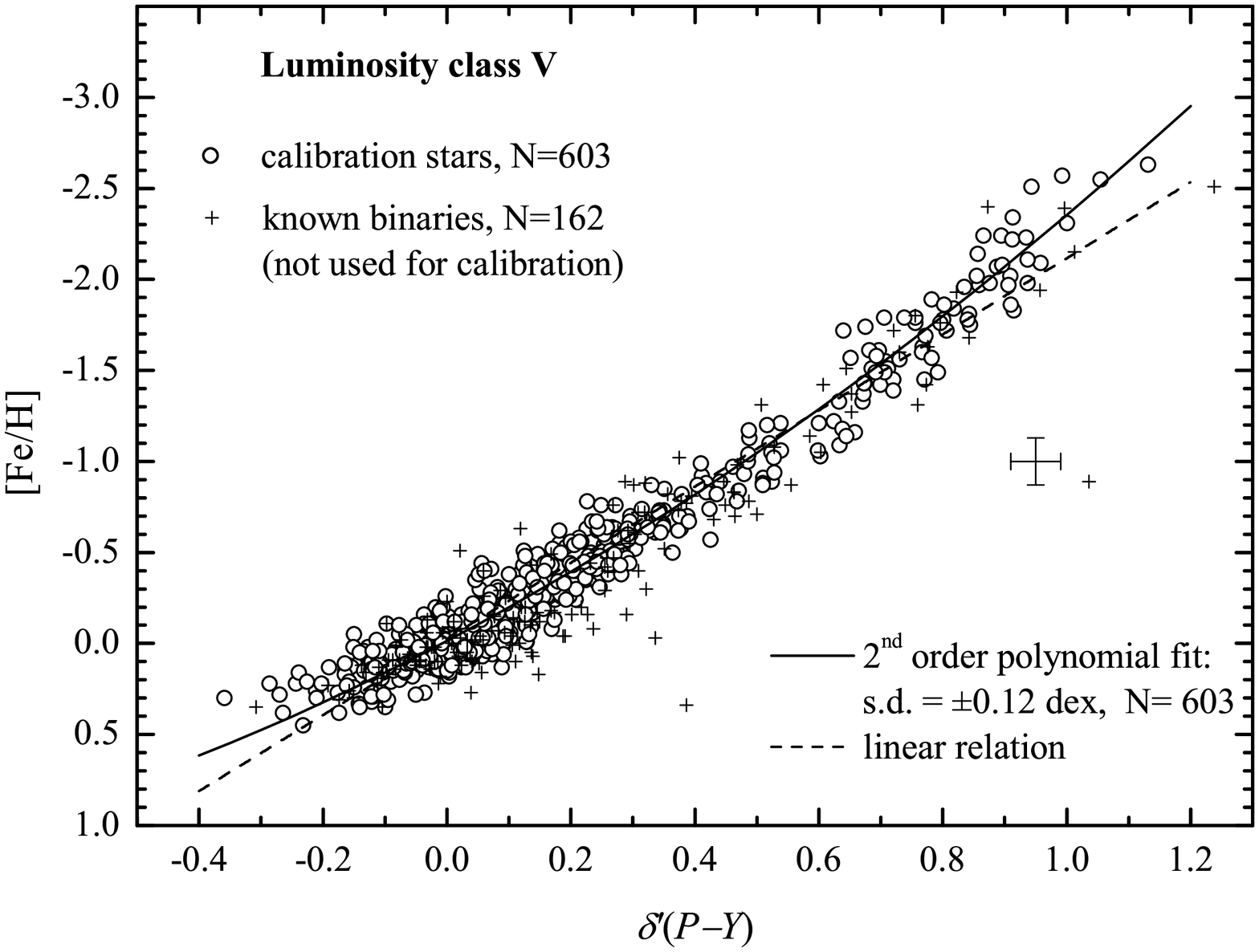,width=106mm,angle=0,clip=}}
\vskip0.5mm \captionb{5b}{$\delta^\prime(P$--$Y)$ vs. [Fe/H]
relation for calibrating dwarf stars with metallicities from
high-dispersion spectroscopy.}
\end{figure}

\begin{figure}[!th]
\centerline{\psfig{figure=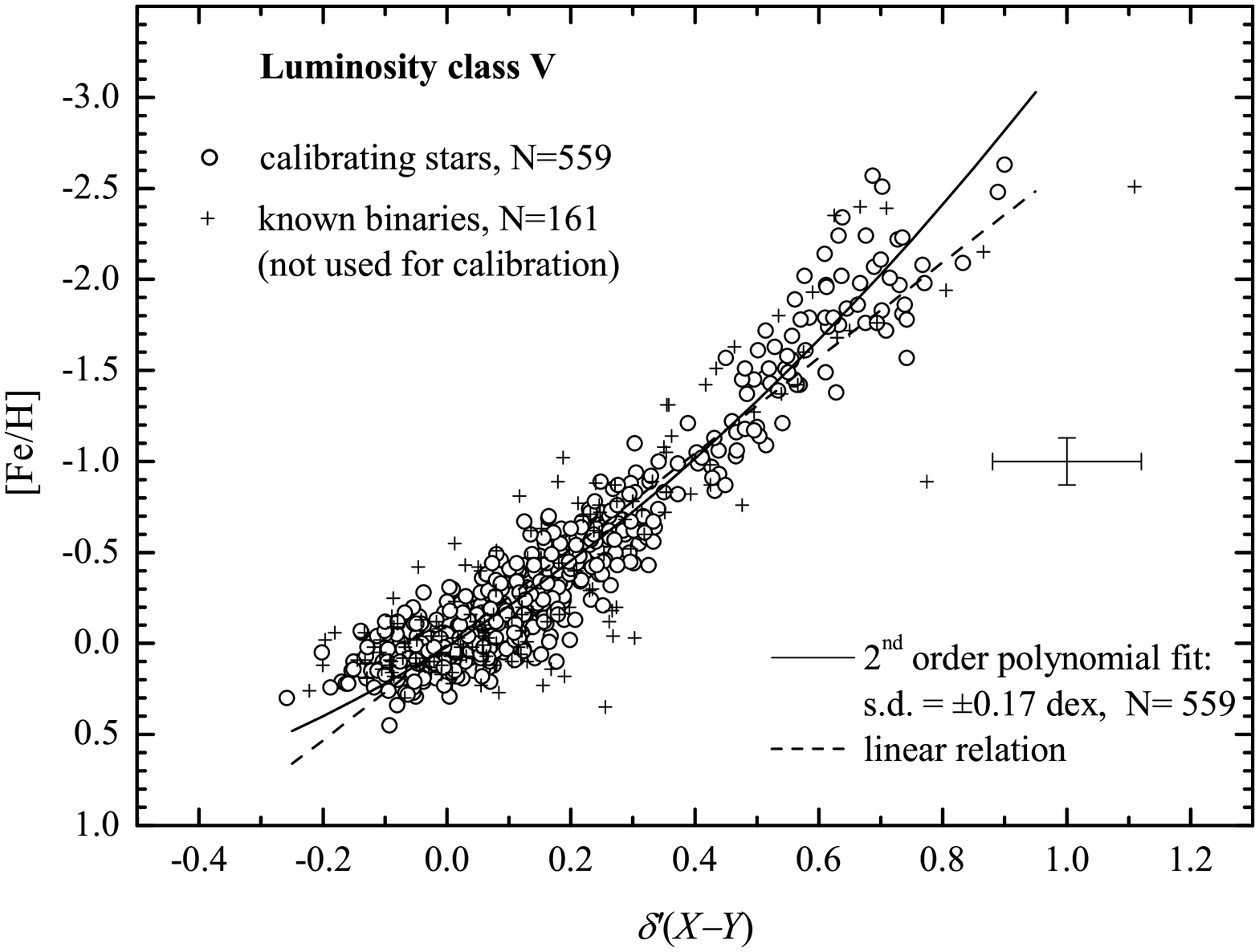,width=106mm,angle=0,clip=}}
\vskip0.5mm \captionb{5c}{$\delta^\prime(X$--$Y)$ vs. [Fe/H]
relation for calibrating dwarf stars with metallicities from
high-dispersion spectroscopy.}
\end{figure}

From the spread of points denoting subdwarfs it is clearly seen that
the maximal deblanketing for the color index $P$--$Y$ is about twice
as large as that for indices $P$--$X$ and $X$--$Y$ (notice that in
the figures the axes scales also differ twice, but in the opposite
sense). The $(CI)_{\rm max}$ lines defined in this paper (thin
continuous line in Figures 4\,a,\,b,\,c) are generally close to
those adopted in BS83 (thin dashed line in Figures 4\,b,c; for
$P$--$X$, BS83 gave no calibration), but extend significantly the
spectral range covered to later types. However, due to the very few
subdwarfs of late K and M types in the sample, the extension of our
empirical $(CI)_{\rm max}$ line to the M-type region should be
considered as very provisional. Despite the fact that the $(CI)_{\rm
max}$ line was constructed in such a way that it should
approximately represent a $-3.0$ dex metallicity, its position in
either of the three diagrams differs significantly from the $-3.0$
dex model curve. It should be noted in this context that the
existence of large differences between observed and synthetic {\it
Vilnius} colors for Kurucz models was earlier shown by Strai\v{z}ys
et al. (2002). At the early-spectral-type end (A8--F0) of the
two-color diagrams, where the lack of subdwarfs in the sample
prevented extension of the $(CI)_{\rm max}$ line into this region,
this line was extrapolated to the bluest point of the $-3.0$ dex
model curve.


In Figures 5\,a,b,c we plot for the calibrating dwarfs the
quantities $\delta^\prime(P$--$X)$, $\delta^\prime (P$--$Y)$ and
$\delta^\prime (X$--$Y)$ versus [Fe/H] from high-dispersion
spectroscopy. The error bars shown in each diagram are estimates of
the average standard deviation in the spread in [Fe/H] from multiple
determinations ($\pm$0.13 dex) and the error of $\delta^\prime(CI)$,
produced by uncertainties in photometry, interstellar-reddening
corrections and two-color relations. As it was mentioned in \S\,2,
known binaries and variable stars were excluded from the
calibrations. In Figures 5\,a,b,c we plotted these stars (marked by
thin crosses) just to illustrate the scatter contributed by the
effects of binarity. A few binary stars deviating very significantly
on the lower right of the relation are also known as variables of RS
CVn and other types. In general, a large fraction of spectroscopic
binaries still fall in the region of calibrating stars.

To obtain the equations of $\delta^\prime(CI)$,\,[Fe/H] relations, a
second-order polynomial was fitted via a least-squares fitting
routine, although a major part of the relation, as can be seen in
Figures 5\,a,b,c, is actually almost linear. Only near both ends of
the metallicity range, [Fe/H]$<-1.5$ and [Fe/H]$>0$, a departure
from linearity is obvious, which may indicate a decreased
metallicity sensitivity of the color indices. For the metal-rich end
of the relation there might also be some other reasons. Polynomial
coefficients of the relations and their errors are given in Table~1.

A comparison of Figures 5\,a,b,c (see also Table 1) indicates that
the calibration of color indices $P$--$X$ and $P$--$Y$ is more
accurate than that of $X$--$Y$. The rms deviations of the standard
[Fe/H] values from the fits are $\pm0.17$ dex for $X$--$Y$,
$\pm0.15$ dex for $P$--$X$ and only $\pm0.12$ dex for $P$--$Y$. This
result confirms the conclusion drawn from the earlier investigation
by Bartkevi{\v c}ius \& Strai\v{z}ys (1970\,a) that the color index
$P$--$Y$ shows the largest deblanketing effect and is most suitable
for estimating metallicity of subdwarf stars.

As a final estimate of photometric metallicity of dwarfs and
subdwarfs, we recommend to use primarily the result based on
$P$--$Y$ or to use an average over the results from $P$--$Y$ and
$P$--$X$, weighted in favor of $P$--$Y$. The color index $X$--$Y$
can be used as an additional means of lower weight. The calibration
is applicable to dwarf stars of spectral types $\sim$A8 to M1
($Y$--$V$\,=\,0.30--1.02, or $Y$--$S$\,=\,0.55--2.22) and the
metallicity range $-2.8<$\,[Fe/H]$<+0.5$. The accuracy of the
calibration and a comparison with BS83 will be discussed in \S\,3.4.

\vskip2mm

\subsectionb{3.3}{Giant stars}

The same procedure, as described in the previous subsection, we
repeated for giant stars. The three two-color diagrams, on which our
calibrations were based, are displayed in Figures 6\,a,b,c. The
diagrams contain the same information as Figures 4\,a,b,c.

As in the case of dwarf stars, the empirical [Fe/H]=0 relations were
determined using stars with $+0.10$\,$\geq$\,[Fe/H]\,$\geq$\,$-0.10$
from high-dispersion spectra, plus an additional number stars
without spectroscopic metallicities (in the figures marked by small
thin crosses) to smooth out the very sparse sampling near the ends
of the spectral-type range. The standard deviation around these
relations is $\pm0.02$ mag. A comparison with the canonical
relations for normal chemical composition stars of luminosity class
III, given by Strai\v{z}ys (1992; Table 68), revealed no significant
differences.

Model curves from Kurucz's grids for [Fe/H]=0, again, do not
satisfactorily match the empirical relations, except for the
$(P$--$Y$,\,$Y$--$S)$ diagram where both curves, theoretical and
empirical, show a remarkable coincidence for $Y$--$S<1.4$. The model
[Fe/H]=--3.0 line (dash-dot-dot) follows quite closely the empirical
$(CI)_{\rm max}$ line (i.e., ``--3 dex isoline") in the
$(P$--$X$,\,$Y$--$V)$ diagram, as well as does over the range of
spectral types earlier than K3 in the other two diagrams. Our
$(CI)_{\rm max}$ lines are very close to those defined in BS83, but
are extended to include earlier spectral types. As in the case of
dwarf stars, the $(CI)_{\rm max}$ line at the early-spectral-type
end was forced to fit the $-3.0$ dex model curve.

\begin{figure}[!th]
\centerline{\psfig{figure=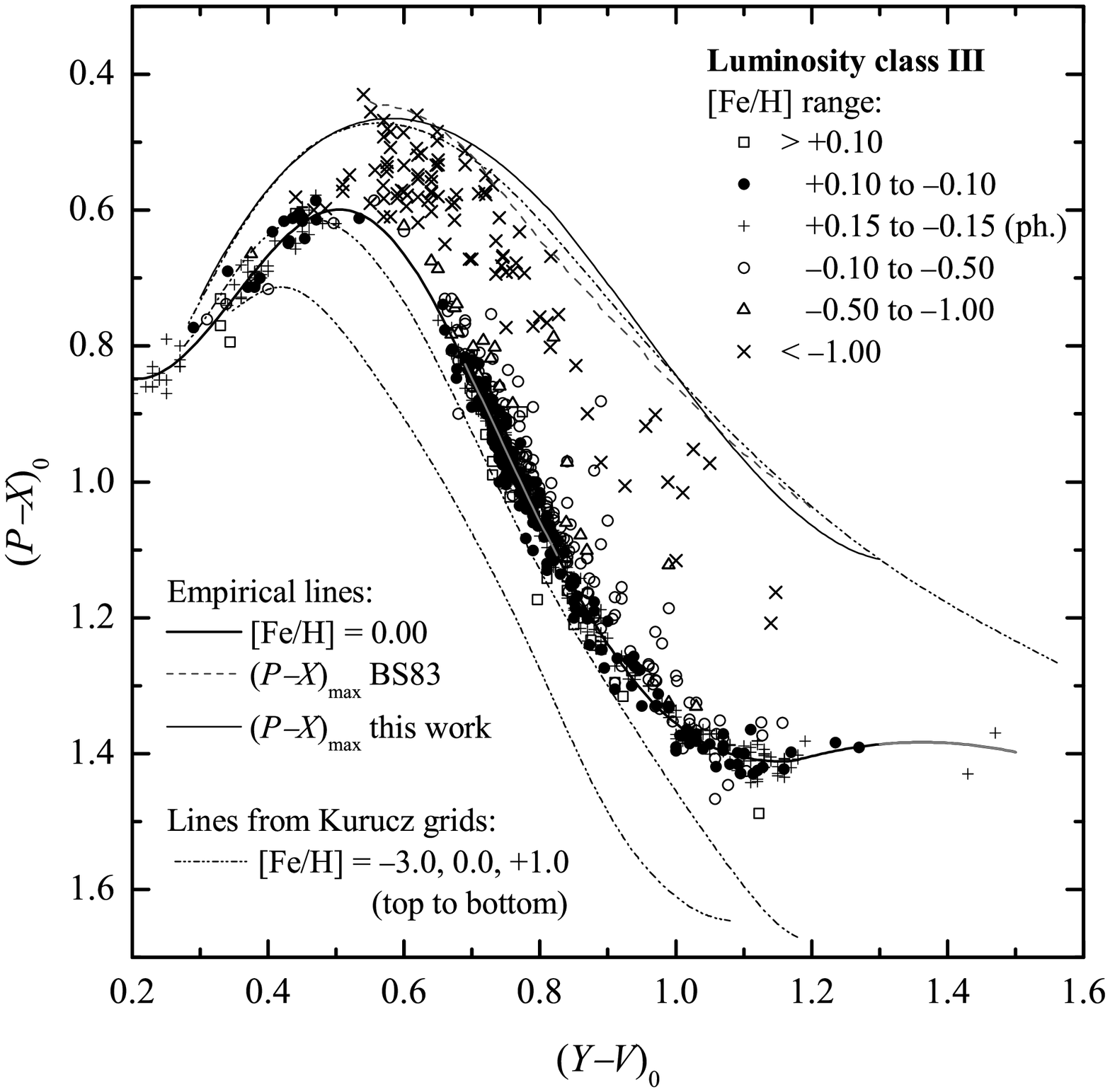,width=85mm,angle=0,clip=}}
\vskip0.5mm \captionb{6a}{$(P$--$X$,\,$Y$--$V)$ diagram for giant
stars. Symbols the same as in Figs.\,4\,a,b.}
\end{figure}
\begin{figure}[!bh]
\centerline{\psfig{figure=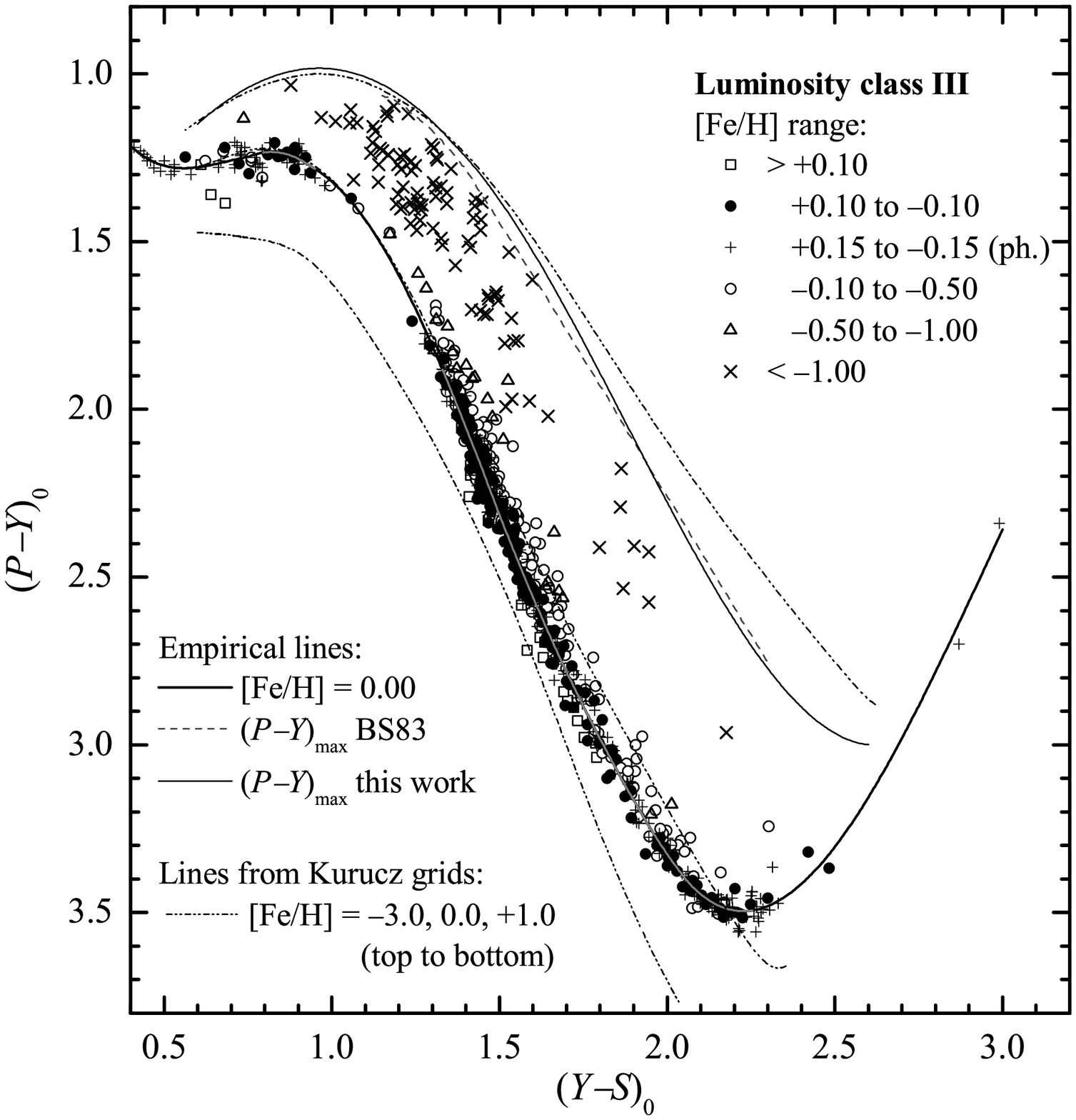,width=85mm,angle=0,clip=}}
\vskip0.5mm \captionb{6b}{$(P$--$Y$,\,$Y$--$S)$ diagram for giant
stars. Symbols the same as in Figs.\,4\,a,b.}
\end{figure}

\begin{figure}[!th]
\centerline{\psfig{figure=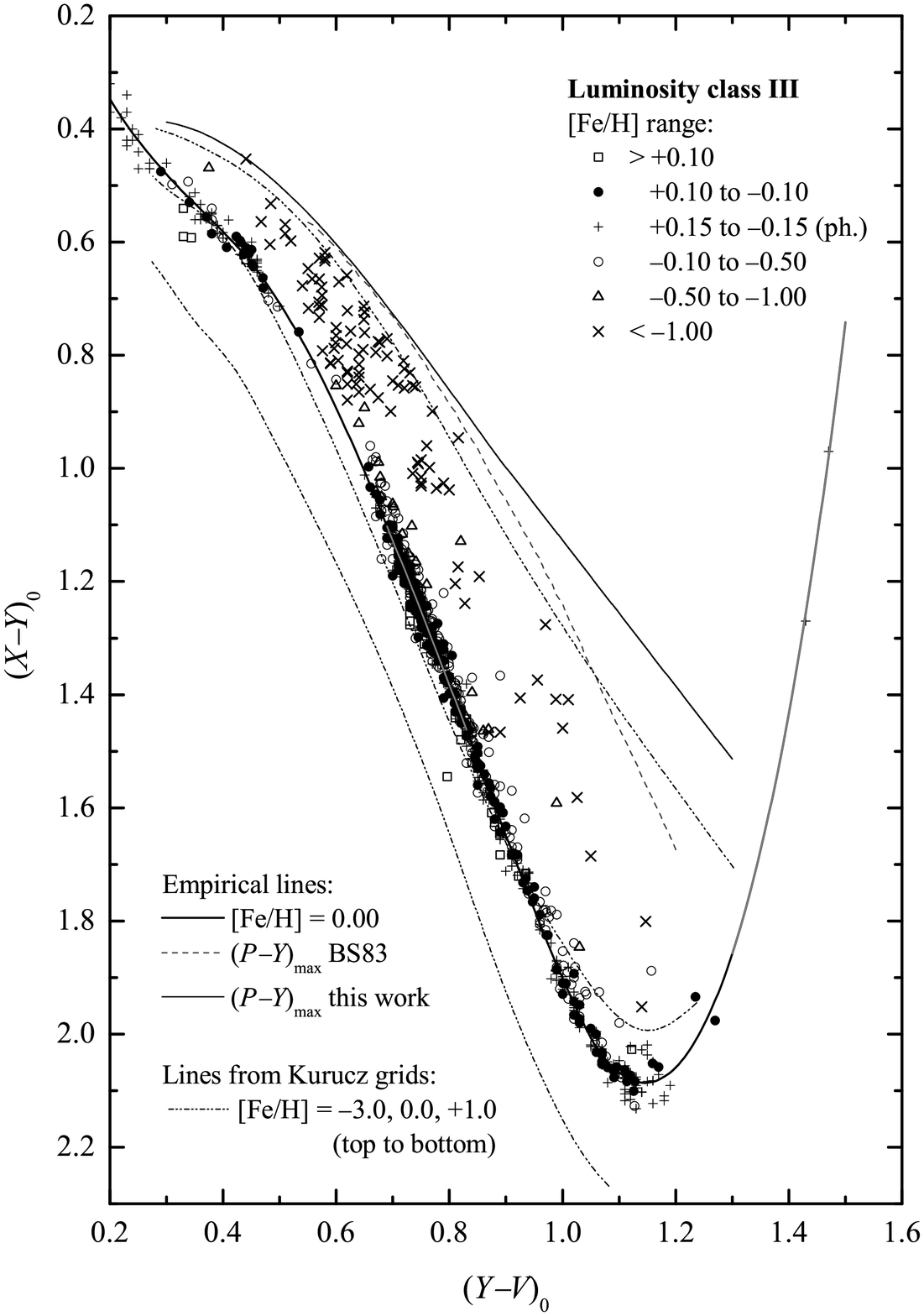,width=85mm,angle=0,clip=}}
\vskip0.5mm \captionb{6c} {$(X$--$Y$,\,$Y$--$V)$ diagram for giant
stars. Symbols as in Figs.\,4\,a,b.}
\end{figure}

The calibrations of $\delta^\prime(CI)$ are presented in Figures
7\,a,b,c. The first feature to notice in these figures is the
absence of calibrating stars in the region around [Fe/H]\,=\,$-1$
dex. It seems likely that this gap, also obvious in histograms of
Figure 2, or, in the two-color plots in Figure 6, reflects a paucity
of stars around the metallicity separating disk and halo populations
rather than a sample bias (this observational fact has long been
noted in the literature, see, e.g. Marsakov \& Suchkov 1977). As in
similar plots for the dwarf stars, known and suspected binaries
(spectroscopic, astrometric, radial-velocity variables) are also
plotted for comparison. In this context, it is interesting to note
that the points denoting binary stars exhibit much less scatter in
the case of $\delta^\prime(P$--$Y)$ (Figure 7\,b) compared to that
for $\delta^\prime(P$--$X)$ and $\delta^\prime(X$--$Y)$ (Figures
7\,a,c).

In all three plots we have almost a linear dependence of
$\delta^\prime(CI)$ on [Fe/H], especially in the case of
$\delta^\prime(X$--$Y)$,\,[Fe/H] relation. However, a 2nd-order
polynomial fit gives a somewhat better fit to the data (the
polynomial coefficients and their errors are given in Table~1). At
low metallicities, the straight line in Figures 7\,a,b,c declines up
very slightly, what indicates that reduction in metal content down
to nearly $-3$ dex does not significantly affect the colors of giant
stars, or, at least, affects the colors less than those of
metal-poor subdwarfs.

\begin{figure}[!th]
\centerline{\psfig{figure=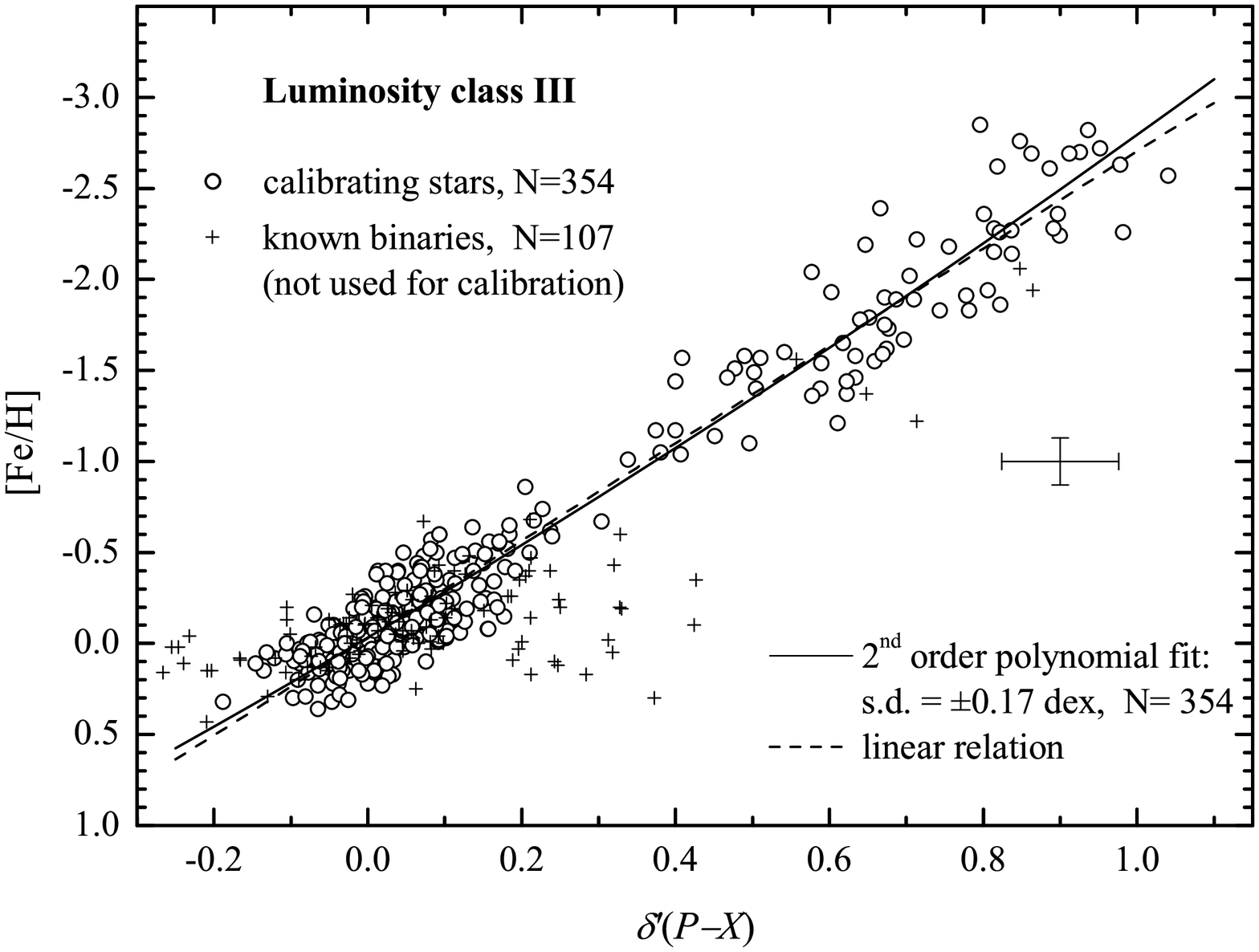,width=106mm,angle=0,clip=}}
\vskip0.5mm \captionb{7a}{$\delta^\prime(P$--$X)$ vs. [Fe/H]
relation for calibrating giant stars with metallicities from
high-dispersion spectroscopy. The solid line is a second-order
polynomial fit and the dashed line is a linear relation. Known
spectroscopic binaries are plotted for comparison.}
\end{figure}
\begin{figure}[!bh]
\centerline{\psfig{figure=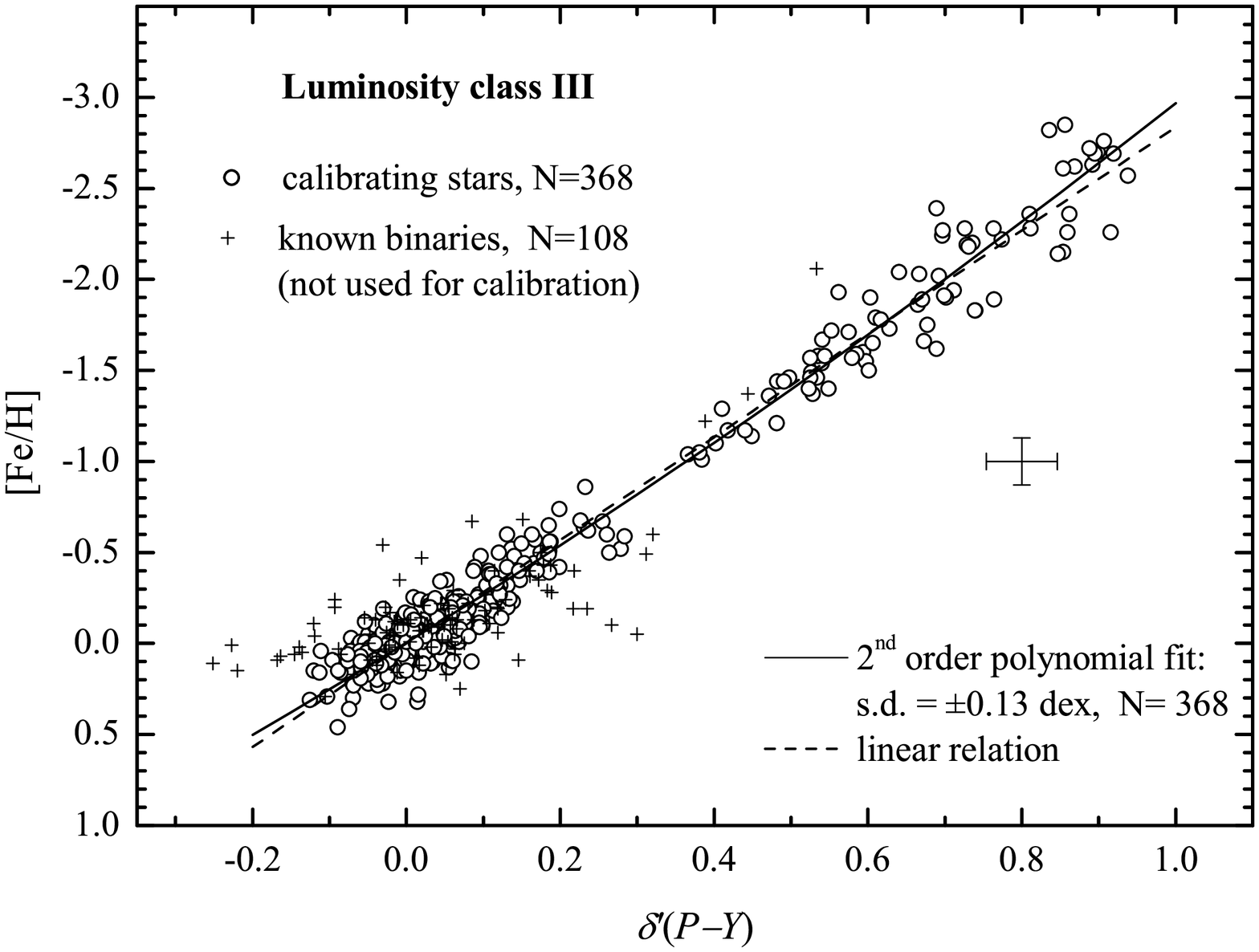,width=106mm,angle=0,clip=}}
\vskip0.5mm \captionb{7b}{$\delta^\prime(P$--$Y)$ vs. [Fe/H]
relation for giant stars with metallicities from high-dispersion
spectroscopy.}
\end{figure}

\begin{figure}[!th]
\centerline{\psfig{figure=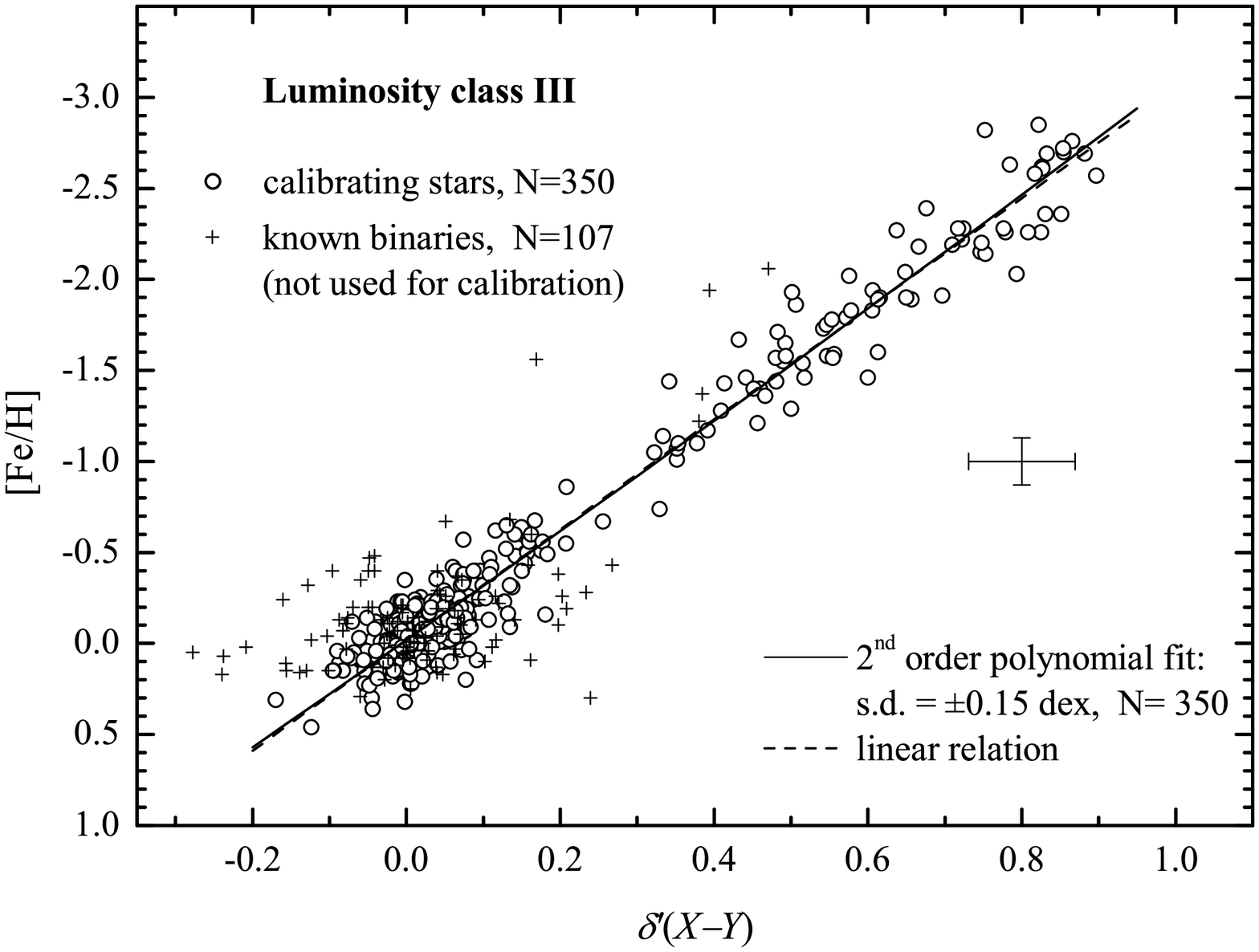,width=106mm,angle=0,clip=}}
\vskip0.5mm \captionb{7c}{$\delta^\prime(X$--$Y)$ vs. [Fe/H]
relation for giant stars with metallicities from high-dispersion
spectroscopy.} \vskip2mm
\end{figure}

The scatter around the defined relations
$\delta^\prime(CI)$,\,[Fe/H] is $\pm0.17$ dex for \hbox{$P$--$X$},
$\pm0.15$ dex for $X$--$Y$ and only $\pm0.13$ dex for $P$--$Y$. As
in the case of dwarf stars, the color index $P$--$Y$ can be most
accurately calibrated in terms of [Fe/H]. A somewhat less accurate
relation was obtained for $P$--$X$ and $X$--$Y$, but the rms
deviations are in general closely the same as in the calibration for
dwarf stars. The new calibration for giants is valid in the
spectral-type range $\sim$A8 to M4 ($Y$--$V$ = 0.30--1.30, or
$Y$--$S$ = 0.60--2.60) and the metallicity range
$-2.8$\,$<$\,[Fe/H]\,$<$\,$+0.5$. The accuracy of the calibration
and a comparison with BS83 will be discussed in the next subsection.


\vskip6mm

\begin{center}
\vbox{\small \centerline{\parbox{96mm}{\baselineskip=9pt {\normbf
Table 1.}{ Polynomial coefficients of the empirical relations
[Fe/H]$=a_0+a_1\delta^\prime(CI)+a_2(\delta^\prime(CI))^2$ for stars
of luminosity class V and III. $N$ is the number of stars used in
the fits, the errors in the last column are standard deviations from
the polynomial fits.}}}}
\medskip
\vbox{ \footnotesize\tabcolsep=3pt
\begin{tabular}{crrrrrr}
\hline
& & & & & & \\[-7pt]
Luminoosity & $\delta^\prime$$(CI)$ & $a_0$~~ & $a_1$~~ & $a_2$~~ & $N$~ & s.d.~ \\
class  &                       &         &         &         & & (dex)\\[-1pt]
  \hline
& & & & & & \\[-6pt]
V & $\delta^\prime$($P$$-$$X$) & $-$0.003 & $-$1.865 & $-$0.792 & 603 & $\pm0.15$ \\
       &                &   $\pm$0.007 &   $\pm$0.048 &   $\pm$0.069 &            & \\
& & & & & & \\[-6pt]
V & $\delta^\prime$($P$$-$$Y$) & $-$0.014 & $-$1.794 & $-$0.546 & 603 & $\pm0.12$ \\
       &                &   $\pm$0.006 &   $\pm$0.038 &   $\pm$0.050 &            & \\
& & & & & & \\[-6pt]
V & $\delta^\prime$($X$$-$$Y$) & 0.018 & $-$2.132 & $-$1.131 & 559 & $\pm0.17$ \\
       &                &   $\pm$0.009 &   $\pm$0.074 &   $\pm$0.122 &            & \\
& & & & & & \\[-6pt]
\hline
 & & & & & &\\[-6pt]
III & $\delta^\prime$($P$$-$$X$) & $-$0.032 & $-$2.497 & $-$0.264 & 354 & $\pm0.17$ \\
       &                &   $\pm$0.011 &   $\pm$0.110 &   $\pm$0.139 &            & \\
& & & & & & \\[-6pt]
III & $\delta^\prime$($P$$-$$Y$) & $-$0.004 & $-$2.603 & $-$0.360 & 368 & $\pm0.13$ \\
       &                &   $\pm$0.008 &   $\pm$0.083 &   $\pm$0.110 &            & \\
& & & & & & \\[-6pt]
III & $\delta^\prime$($X$$-$$Y$) & $-$0.019 & $-$2.968 & $-$0.113 & 350 & $\pm0.15$\\
       &                &   $\pm$0.009 &   $\pm$0.105 &   $\pm$0.144 &            & \\
& & & & & & \\[-6pt]
\hline
\end{tabular}
}
\end{center}

\newpage

\subsectionb{3.4}{Comparison to BS\,83 and error analysis}

In Figures 8 and 9 we compare the differences between [Fe/H] from
high-dispersion spectroscopy and [Fe/H] determined from old (BS83)
and new calibrations for the stars in our sample, plotted as
functions of metallicity (left-hand panels) and the
temperature-dependent color indices $Y$--$V$ or $Y$--$S$ (right-hand
panels). The residuals $\Delta$[Fe/H] plotted are in the sense
``spectroscopic minus photometric".

Consider first Figure 8, in which metallicities for stars of
luminosity class V are plotted (in BS83, however, $P$--$X$
calibration for dwarfs was not performed). The different numbers of
stars, indicated in the upper (BS83) and lower (our calibration)
panels, reflect the fact that BS83 calibration covers a smaller
spectral range. The figure shows that, in addition to a larger
spread of residuals ($\Delta$[Fe/H]$\sim \pm$0.20 dex), the old
calibration of $P$--$Y$ and $X$--$Y$ systematically overestimates
(by $\sim$0.1 dex, on average) the metallicities over most of the
[Fe/H] range. There is also a systematic trend with metallicity in
the results from BS83 in the range $+0.5$\,$>$\,[Fe/H]\,$>$\,$-1.0$.
By application of the new calibration this systematic trend
disappears. Figure 8 is enough to convince one that our revised
calibration for dwarfs and subdwarfs has a distinct advantage over
that of BS83.

It should be noted, however, that the spread of the residuals
between the values from high-dispersion spectroscopy and our
photometric values is not the same over the entire range of
metallicities. For stars with [Fe/H]\,$>$\,$-1$, the standard
deviation is the same or somewhat smaller than that estimated from
the whole sample of calibrating dwarfs and given in Figure~8 or
Table 1. Considering only subdwarfs with [Fe/H]\,$<$\,$-1$ (87
stars), we estimate rms errors of $\pm 0.15$ and $\pm 0.17$ dex when
using calibrations of $P$--$Y$ and $P$--$X$, respectively, while in
the case of $X$--$Y$ the rms error becomes as large $\pm 0.24$ dex
(using BS83 calibration, $\pm 0.26$ dex).

As an independent check on accuracy of our calibration for dwarf
stars of chemical composition close to solar, including an estimate
of the photometric metallicity zero-point error, we applied the new
calibration for the Hyades dwarfs, for which {\it Vilnius}
photometry was available. Considering a set 54 stars (excluding
known spectroscopic binaries), we find an average of
[Fe/H]\,$=$\,$+0.10\pm0.01$ with a dispersion of $\pm0.11$ dex from
the $P$--$Y$ colors and
[Fe/H]\,$=$\,$+0.13\pm0.02$\,(s.d.\,$=\pm0.15$) from $P$--$X$. The
less accurate calibration of $X$--$Y$ gives for the Hyades an
average of $+0.10\pm0.03$\,(s.d.\,$=\pm0.21$) dex. The average
photometric metallicity values are consistent with spectroscopic
estimates in the literature (e.g., Paulson et al. (2003) from a
differential abundance analysis of Hyades F--K dwarfs give
[Fe/H]\,$=$\,$+0.13\pm0.01$, s.d.\,$=\pm0.05$). We can therefore
conclude that our new calibration seems to reproduce [Fe/H] values
for dwarfs of near-solar metallicity without a zero-point offset.

\begin{figure}[!t]
\centerline{\psfig{figure=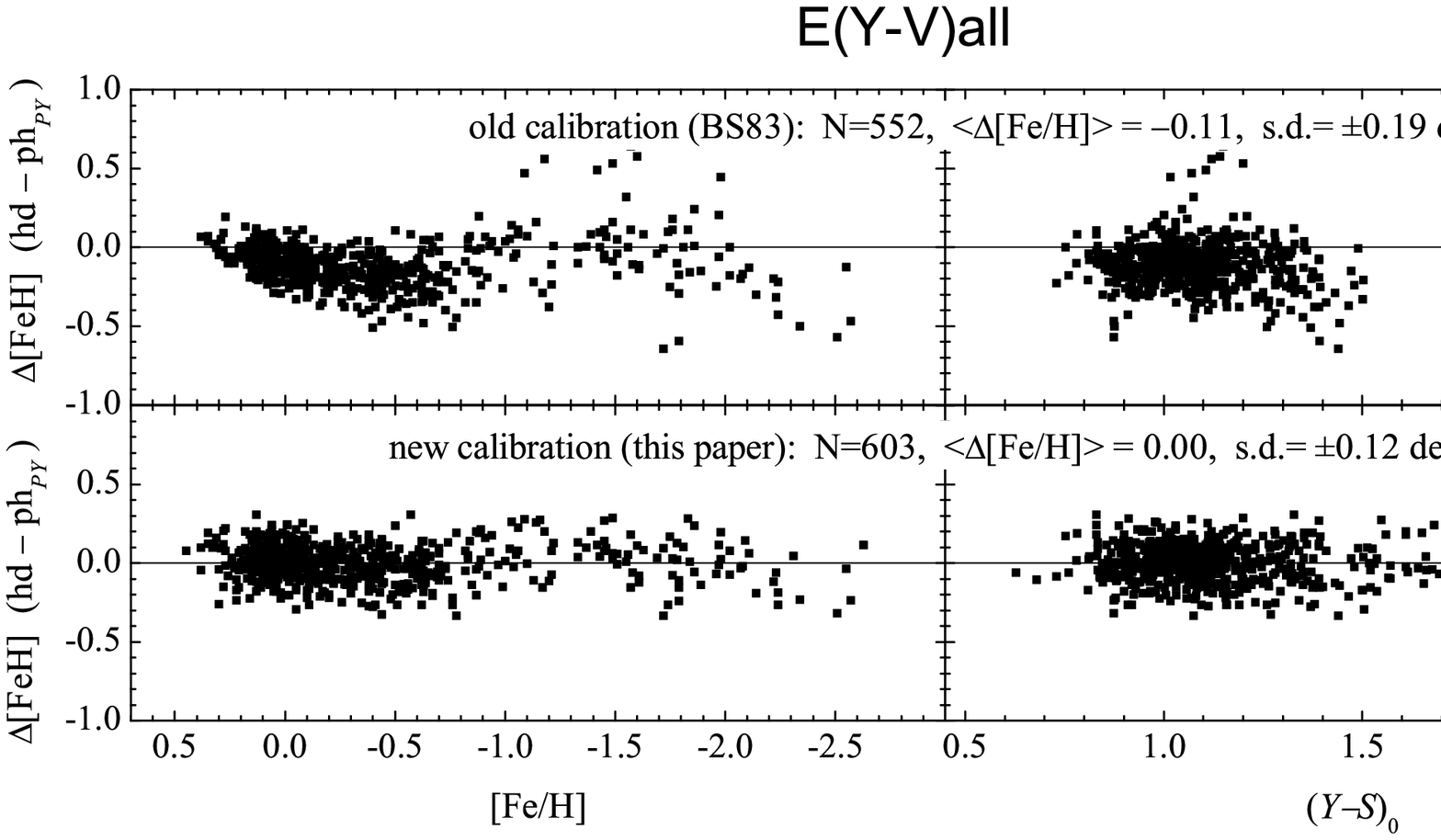,width=124mm,angle=0,clip=}}
\vskip3mm
\centerline{\psfig{figure=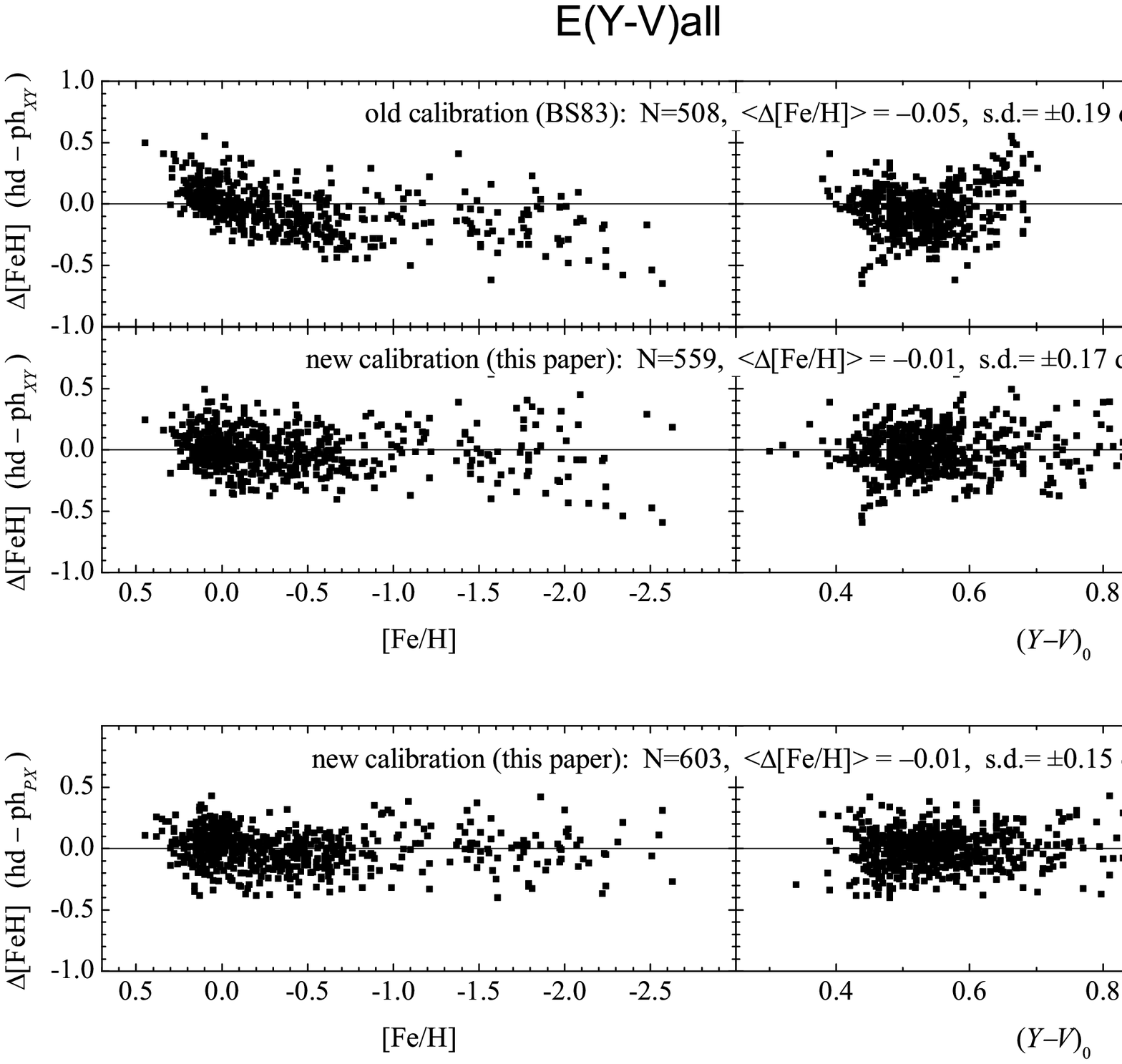,width=124mm,angle=0,clip=}}
\vskip1mm \captionb{8}{Differences between [Fe/H] from
high-dispersion spectroscopy and [Fe/H] from old (BS83) and new
calibrations of {\em Vilnius} photometry for the same set of dwarf
stars, shown as functions of [Fe/H] (left-had panels) and
temperature-dependent color indices (right-hand panels). The number
of stars compared, mean residuals and standard deviations are given
on the top of each graph.}
\end{figure}

\begin{figure}[!t]
\centerline{\psfig{figure=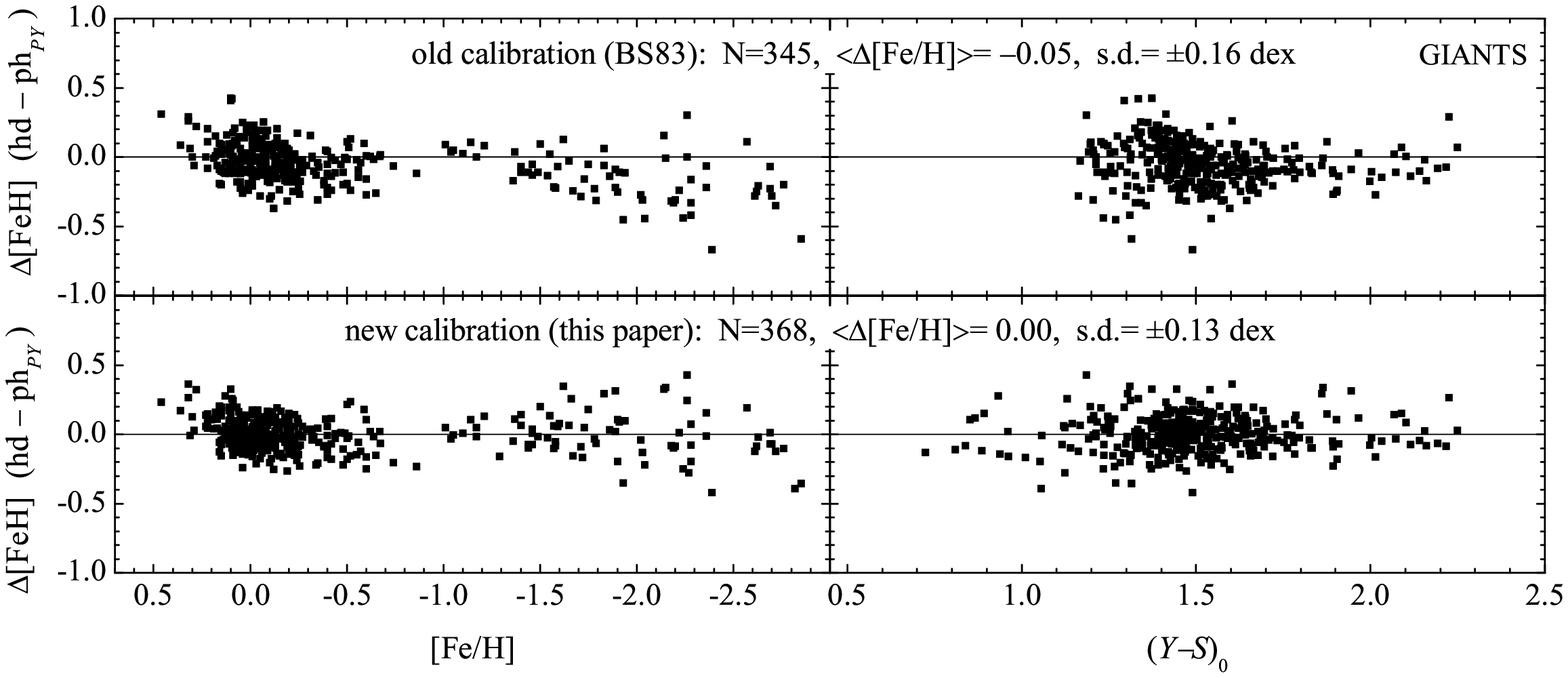,width=124mm,angle=0,clip=}}
\vskip3mm
\centerline{\psfig{figure=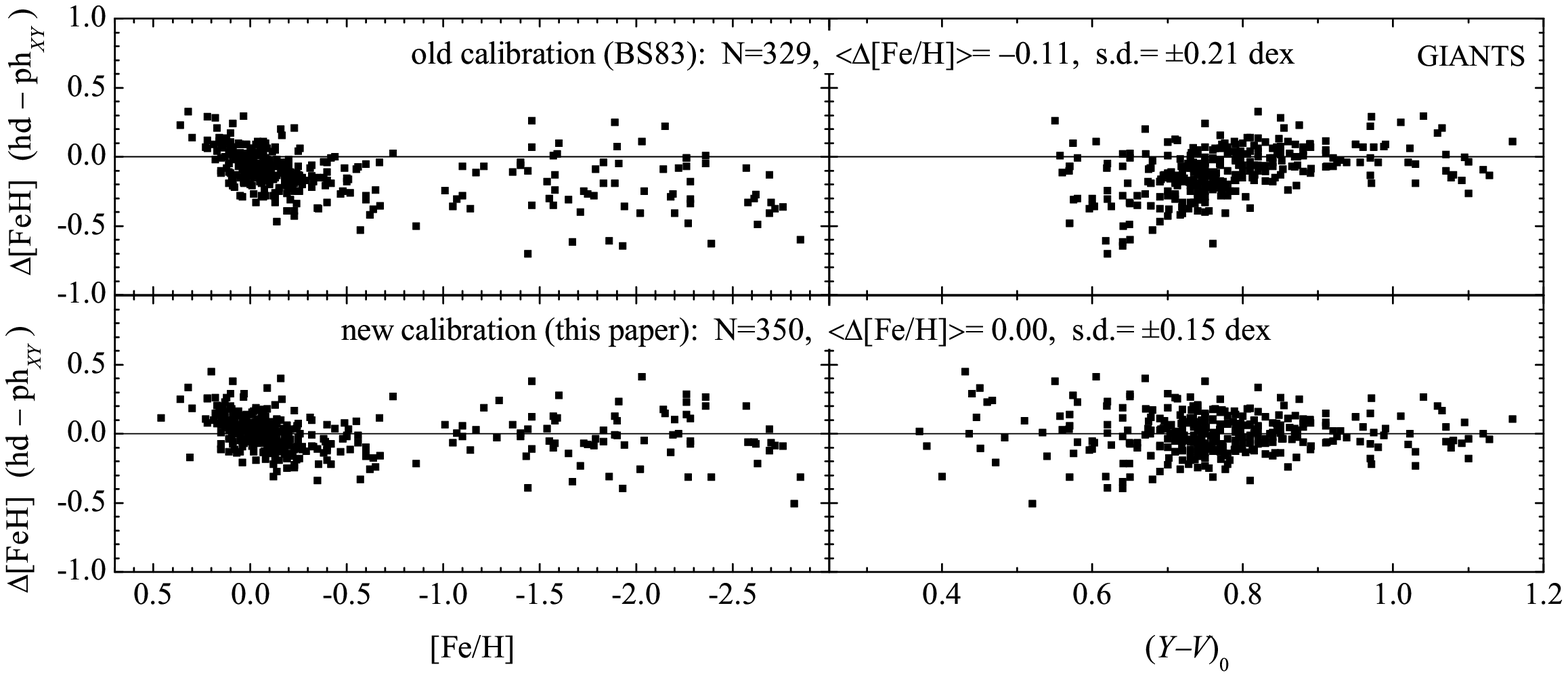,width=124mm,angle=0,clip=}}
\vskip3mm
\centerline{\psfig{figure=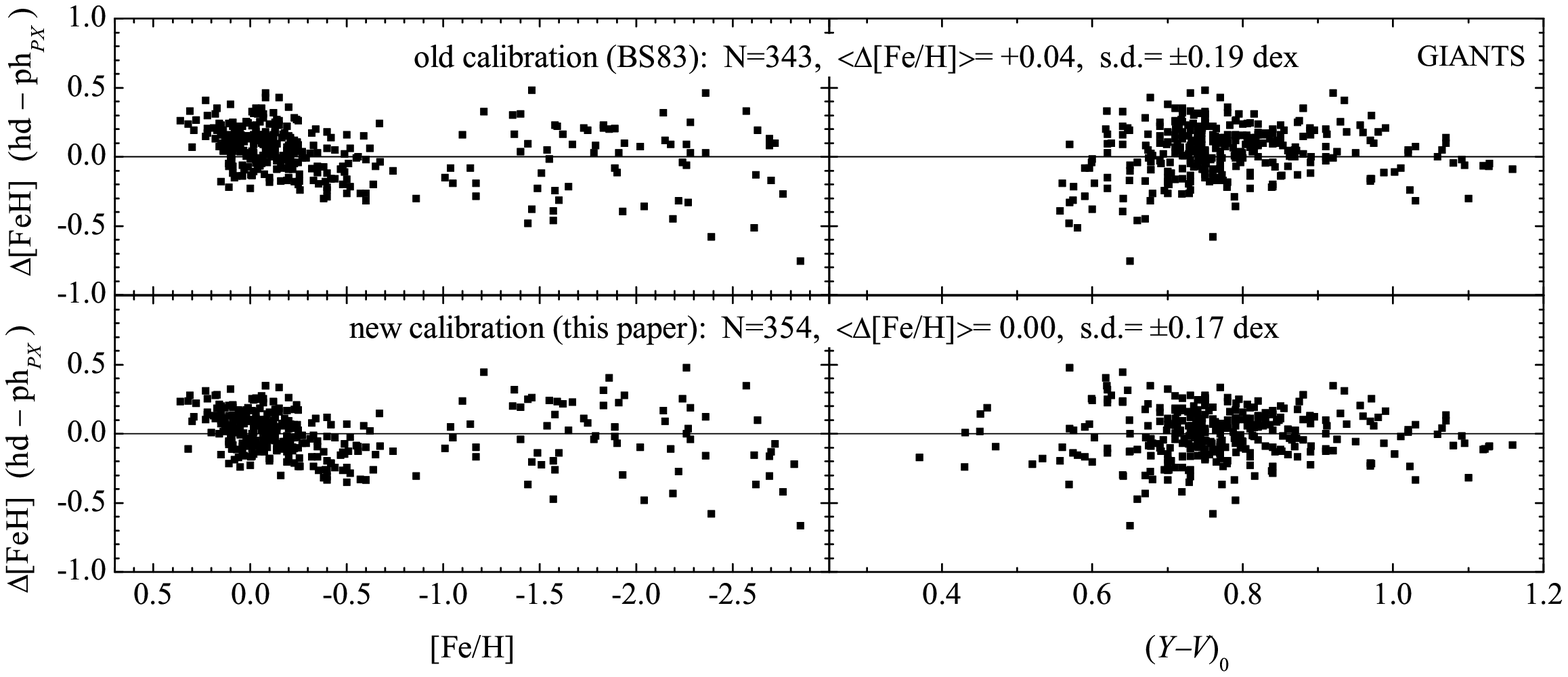,width=124mm,angle=0,clip=}}
\vskip1mm \captionb{9}{Differences between [Fe/H] from
high-dispersion spectroscopy and [Fe/H] from old (BS83) and new
calibrations of {\em Vilnius} photometry for the same set of giant
stars, shown as functions of [Fe/H] (left-had panels) and
temperature-dependent color indices (right-hand panels). The number
of stars compared, mean residuals and standard deviations are given
on the top of each graph.}
\end{figure}

Let us now consider Figure 9, where comparisons are made for stars
of luminosity class III. The most important feature to notice in
this figure is the slope in the region
$+0.5$\,$>$\,[Fe/H]\,$>$\,$-1.0$, most pronounced in the
distribution of points from $P$--$X$ and $X$--$Y$ calibrations. It
shows that both photometric calibrations, old and new,
systematically underestimate metallicities for giant stars with
[Fe/H]$>0.0$ and overestimates in the range
$-0.2$\,$>$\,[Fe/H]\,$>$\,$-1.0$. It is not easy to find an
explanation for this discrepancy between spectroscopic and
photometric [Fe/H]. We can suggest that at least one of the sources
lies in the dependence of the color indices $P$--$X$ and $X$--$Y$ on
C and N abundances. The {\it Vilnius} passbands $P$, $X$ and $Y$
include, in the case of late-type giants of chemical composition
close to solar, rather many C$_2$, CH and CN molecular bands.
Therefore, the corresponding color indices can predominantly be
affected by the combined effect of abundances of Fe and C (and its
compounds) rather than by the pure Fe abundance against which they
were calibrated.

A comparison with BS83 shows that the new calibration of the color
index \hbox{$P$--$X$} shows no appreciable improvement however. In
the case of $P$--$Y$ and \hbox{$X$--$Y$}, our calibration has
removed the systematic differences which are apparent in the `BS83'
metallicity values for metal-deficient giants with
[Fe/H]\,$<$\,$-1$. For the latter stars, the new calibration of
$P$--$Y$ and $X$--$Y$ reproduces [Fe/H] values to within $\pm 0.17$
and $\pm 0.18$ dex, respectively, while for giants with
[Fe/H]\,$>$\,$-1$ the rms differences from spectroscopic values
become as small as $\pm 0.12$ and $\pm 0.13$ dex, respectively. In
this metallicity range, $P$--$X$ gives also satisfactory results,
$\pm 0.15$ dex, but for [Fe/H]\,$<$\,$-1$, our calibration of
$P$--$X$ can reproduce [Fe/H] values only to within $\pm 0.25$ dex
(using old calibration, to $\pm 0.27$ dex).

\sectionb{4}{CONCLUDING REMARKS}

For deriving metallicities from {\it Vilnius} photometry the revised
empirical calibration presented here should be used, rather than the
previous empirical calibration by BS83. The new calibration, based
on the color indices $P$--$X$, $P$--$Y$ and $X$--$Y$, is applicable
to dwarf and giant stars of spectral types F--M in the metallicity
range $-2.8$\,$<$\,[Fe/H]\,$<$\,$+0.5$. Both for dwarfs and giants,
the color index $P$--$Y$ should be regarded as the most accurate and
sensitive metallicity indicator. Using $P$--$Y$, [Fe/H] values can
be determined with an accuracy of $\pm$0.12 dex for stars of solar
and mildly sub-solar metallicity and $\pm0.17$ dex for stars with
[Fe/H]\,$<$\,$-1$. In the latter metallicity range, a more reliable
estimate of photometric metallicities would be an average over the
results from $P$--$X$ and $P$--$Y$ for dwarf stars and from $P$--$Y$
and $X$--$Y$ for giant stars, weighted, in either case, in favor of
$P$--$Y$. For [Fe/H]\,$>$\,$-1$, all of the three color indices can
be used, giving them proper weights.

We made no attempt to calibrate {\it Vilnius} color indices for
stars of luminosity class IV. Subgiants in our initial sample were
not numerous (a total of 157 stars), and among them only eight stars
have [Fe/H]\,$<$\,$-1$. Instead, we would recommend for deriving
their metallicities the so-called  `comparison' method, described in
detail and applied by Bartkevi{\v c}ius \& Lazauskait{\. e} (1996,
1997). The principle of the method is to find from an extended data
bank a set of standard stars having closely the same intrinsic color
indices (or the same reddening-free parameters $Q$) as a star of
interest and, then, to ascribe to the latter the values of [Fe/H]
(and other physical parameters), averaged over the extracted set
standard stars. To check the validity of this approach, we compiled
a data bank of 1666 stars from our sample, all with standard
spectroscopic values of [Fe/H], and determined by the `comparison'
method their metallicities. The differences from the standard [Fe/H]
values were found to be $\pm0.16$ dex for dwarfs, $\pm0.19$ dex for
subgiants and $\pm0.25$ dex for subdwarfs and all giants. Although
of lower accuracy than the new calibration, the `comparison' method
has two advantages. One is that it does not necessarily require
dereddening of color indices, once the reddening-free parameters $Q$
can be used instead of intrinsic colors. Another advantage is that
it allows us to derive photometric metallicity of binary stars if a
data bank of standards contains a sufficient variety of binaries.
For 272 binary stars in our sample, for example, the metallicities
derived by the `comparison' method have a rms difference from
spectroscopic values of only $\pm0.20$ dex.

A list of 971 calibrating stars with their parameters, used in the
present work, can be supplied by authors on request.

\thanks{This research was supported by the Research Council of
Lithuania under grant No. MIP-132/2010. The data search was greatly
aided through the use of the SIMBAD database, operated at the CDS,
Strasbourg.}

\References

\refb Afsar~M., Sneden~C., For~B.-Q. 2012, AJ, 144, 20

\refb Barta{\v s}i{\= u}t{\. e} S., Janusz~R., Boyle~R.\,P.,
Philip~A.\,G.\,D. 2011, Baltic Astronomy, 20, 27

\refb Bartkevi{\v c}ius~A., Lazauskait{\. e}~R. 1996, Baltic
Astronomy, 5, 1

\refb Bartkevi{\v c}ius~A., Lazauskait{\. e}~R. 1997, Baltic
Astronomy, 6, 499

\refb Bartkevi{\v c}ius~A., Sperauskas~J. 1983, Bull. Vilnius Obs.,
63, 3 (BS83)

\refb Bartkevi{\v c}ius~A., Strai\v{z}ys~V. 1970a, Bull. Vilnius
Obs., 28, 33

\refb Bartkevi{\v c}ius~A., Strai\v{z}ys~V. 1970b, Bull. Vilnius
Obs., 30, 16

\refb Bartkevi{\v c}ius~A., Sviderskien{\. e}~Z. 1981, Bull. Vilnius
Obs., 58, 54

\refb Bond~H.\,E. 1980, ApJ, 44, 517

\refb Cayrel de Strobel~G., Soubiran~C., Friel~E.\,D., Ralite~N.,
Fran{\c c}ois~P. 1997, A\&AS, 124, 299

\refb Cayrel de Strobel~G., Soubiran~C., Ralite~N. 2001, A\&A, 373,
159

\refb Famaey~B., Jorissen~A., Luri~X., Mayor~M., Udry~S.,
Dejonghe~H., Turon~C. 2005, A\&A, 430, 165

\refb Kazlauskas~A. 2010, {\it General Photometric Catalogue of
Stars Observed in the\break Vilnius System} (in preparation),
private communication

\refb Kurilien{\. e}~G., S{\= u}d{\v z}ius~J. 1974, Bull. Vilnius
Obs., 40, 10

\refb Kurucz R. L. 2001. Grids of model atmospheres (available on
line at\break http://kurucz.harvard.edu/grids.html)

\refb Leeuwen~F., van 2007, A\&A, 474, 653

\refb Liu~Y.\,J., Zhao~G., Shi~J.\,R., Pietrzy{\'n}ski~G., Gieren~W.
2007, MNRAS 382, 553

\refb Marsakov~V.\,A., Suchkov~A.\,A. 1977, AZh, 54, 1232

\refb McWilliam~A. 1990, ApJS, 74, 1075

\refb Nidever~D.\,L., Marcy~G.\,W., Butler~R.\,P., Fischer~D.\,A.,
Vogt~S.\,S. 2002, ApJS, 141, 503

\refb Nordstroem~B., Mayor~M., Andersen~J., Holmberg~J., et al.
2004, A\&A, 418, 989

\refb Paulson~D.\,B., Sneden~C., Cochran~W.\,D. 2003, AJ, 125, 3185

\refb Pourbaix~D., Tokovinin~A.\,A., Batten~A.\,H., et al.
2004--2009, SB9: 9th Catalogue of Spectroscopic Binary Orbits

\refb Samus~N.\,N., Durlevich~O.\,V., Kazarovets~E.\,V.,
Kireeva~N.\,N., Pastukhova~E.\,N., Zharova~A.\,V., et al.
2007--2012, {\it General Catalogue of Variable Stars}\\ (VizieR
On-line Data Catalog: B/gcvs)

\refb Soubiran~C., Le Campion~J.-F., Cayrel de Strobel~G., Caillo~A.
2010, A\&A, 515, A111

\refb Strai\v{z}ys~V. 1992, {\it Multicolor Stellar Photometry},
Pachart Publishing House,\\ Tucson, Arizona; available in pdf format
from\\ http://www.itpa.lt/MulticolorStellarPhotometry/

\refb Strai\v{z}ys~V. Kazlauskas~A. 1993, {\it General Photometric
Catalogue of Stars Observed in the Vilnius System}, Baltic
Astronomy, 2, 1

\refb Strai\v{z}ys~V., Lazauskait{\. e}~R., Valiauga~G. 2002, Baltic
Astronomy, 11, 341

\refb Strai{\v z}ys~V., Boyle~R.\,P., Janusz~R., Laugalys~V.,
Kazlauskas~A. 2013, A\&A, 554, 3

\refb Zdanavi{\v c}ius~J., Vrba~F.\,J., Zdanavi{\v c}ius~K.,
Strai{\v z}ys~V., Boyle~R.\,P. 2011, Baltic Astronomy, 20, 1

\refb Zdanavi{\v c}ius~K., Strai{\v z}ys~V., Zdanavi{\v c}ius~J.,
Chmieliauskait{\. e}~R., Kazlauskas~A. 2012, A\&A, 544, 49

\newpage

\scriptsize
\begin{longtable}{p{0.08\textwidth}p{0.08\textwidth}p{0.08\textwidth}|p{0.08\textwidth}p{0.08\textwidth}p{0.08\textwidth}}

\caption*{{\small{\bf Appendix.} The two-color relations defining
the [Fe/H]=0 isoline, $(CI)_{\rm
n}$, and the isoline of maximal metal-deficiency, $(CI)_{\rm m}$.}}\\[30pt]
\caption*{{\small{\bf Table A1}. ($P$--$Y$, $Y$--$S$) relations for luminosity class V.}}\\[6pt]
\hline

\multicolumn{1}{p{0.08\textwidth}}{\newline$Y$--$S$\newline}&\multicolumn{1}{p{0.08\textwidth}}{\newline($P$--$Y$)\rlap{$_{\rm
n}$}}&\multicolumn{1}{p{0.08\textwidth}|}{\newline($P$--$Y$)\rlap{$_{\rm
m}$}}&\multicolumn{1}{p{0.08\textwidth}}{\newline$Y$--$S$}&\multicolumn{1}{p{0.08\textwidth}}{\newline($P$--$Y$)\rlap{$_{\rm
n}$}}&\multicolumn{1}{p{0.08\textwidth}}{\newline($P$--$Y$)\rlap{$_{\rm
m}$}}
\\
\hline
\endhead
\newline0.55&\newline1.274&\newline1.152&\newline1.61&\newline2.564&\newline2.100\\
0.57&1.271&1.139&1.63&2.599&2.130\\
0.59&1.267&1.125&1.65&2.633&2.159\\
0.61&1.261&1.111&1.67&2.660&2.187\\
0.63&1.254&1.097&1.69&2.685&2.214\\
0.65&1.246&1.083&1.71&2.708&2.241\\
0.67&1.238&1.068&1.73&2.729&2.269\\
0.69&1.230&1.054&1.75&2.747&2.295\\
0.71&1.222&1.040&1.77&2.765&2.320\\
0.73&1.215&1.026&1.79&2.780&2.344\\
0.75&1.209&1.013&1.81&2.793&2.367\\
0.77&1.204&1.000&1.83&2.805&2.391\\
0.79&1.201&0.987&1.85&2.816&2.414\\
0.81&1.199&0.975&1.87&2.825&2.437\\
0.83&1.200&0.965&1.89&2.832&2.458\\
0.85&1.203&0.955&1.91&2.838&2.478\\
0.87&1.208&0.947&1.93&2.842&2.496\\
0.89&1.215&0.939&1.95&2.845&2.513\\
0.91&1.225&0.934&1.97&2.847&2.528\\
0.93&1.238&0.931&1.99&2.848&2.542\\
0.95&1.253&0.928&2.01&2.850&2.554\\
0.97&1.271&0.925&2.03&2.848&2.564\\
0.99&1.292&0.924&2.05&2.846&2.573\\
1.01&1.316&0.924&2.07&2.842&2.581\\
1.03&1.342&0.927&2.09&2.838&2.585\\
1.05&1.372&0.932&2.11&2.834&2.587\\
1.07&1.404&0.940&2.13&2.828&2.588\\
1.09&1.438&0.953&2.15&2.822&2.586\\
1.11&1.475&0.971&2.17&2.816&2.581\\
1.13&1.514&0.999&2.19&2.809&2.574\\
1.15&1.555&1.037&2.21&2.801&2.565\\
1.17&1.599&1.080&2.23&2.793&-\\
1.19&1.644&1.127&2.25&2.784&-\\
1.21&1.691&1.179&2.27&2.775&-\\
1.23&1.739&1.235&2.29&2.766&-\\
1.25&1.788&1.292&2.31&2.756&-\\
1.27&1.837&1.352&2.33&2.746&-\\
1.29&1.888&1.410&2.35&2.736&-\\
1.31&1.936&1.466&2.37&2.725&-\\
1.33&1.984&1.522&2.39&2.714&-\\
1.35&2.031&1.577&2.41&2.703&-\\
1.37&2.077&1.629&2.43&2.692&-\\
1.39&2.122&1.678&2.45&2.680&-\\
1.41&2.165&1.726&2.47&2.669&-\\
1.43&2.208&1.771&2.49&2.657&-\\
1.45&2.250&1.814&2.51&2.645&-\\
1.47&2.292&1.856&2.53&2.633&-\\
1.49&2.333&1.895&2.55&2.620&-\\
1.51&2.374&1.933&2.57&2.608&-\\
1.53&2.413&1.969&2.59&2.596&-\\
1.55&2.452&2.004&2.61&2.583&-\\
1.57&2.490&2.037&2.63&2.571&-\\
1.59&2.528&2.069&2.65&2.558&-\\[3pt]

\hline
\end{longtable}
\pagebreak

\begin{longtable}{p{0.08\textwidth}p{0.08\textwidth}p{0.08\textwidth}|p{0.08\textwidth}p{0.08\textwidth}p{0.08\textwidth}}

\caption*{{\small{\bf Table A2}. ($P$--$Y$, $Y$--$S$) relations for luminosity class III.}}\\[4pt]
\hline
\multicolumn{1}{p{0.08\textwidth}}{\newline$Y$--$S$\newline}&\multicolumn{1}{p{0.08\textwidth}}{\newline($P$--$Y$)\rlap{$_{\rm
n}$}}&\multicolumn{1}{p{0.08\textwidth}|}{\newline($P$--$Y$)\rlap{$_{\rm
m}$}}&\multicolumn{1}{p{0.08\textwidth}}{\newline$Y$--$S$}&\multicolumn{1}{p{0.08\textwidth}}{\newline($P$--$Y$)\rlap{$_{\rm
n}$}}& \multicolumn{1}
{p{0.08\textwidth}}{\newline($P$--$Y$)\rlap{$_{\rm m}$}} \\
[2pt] \hline
\endhead
\newline0.40&\newline1.215&\newline-&\newline1.64&\newline2.652&\newline1.599\\
0.42&1.232&-&1.66&2.697&1.634\\
0.44&1.247&-&1.68&2.740&1.670\\
0.46&1.258&-&1.70&2.783&1.706\\
0.48&1.267&-&1.72&2.824&1.742\\
0.50&1.274&-&1.74&2.865&1.780\\
0.52&1.278&-&1.76&2.905&1.817\\
0.54&1.281&-&1.78&2.944&1.855\\
0.56&1.281&-&1.80&2.982&1.893\\
0.58&1.281&-&1.82&3.020&1.931\\
0.60&1.279&1.150&1.84&3.058&1.970\\
0.62&1.276&1.132&1.86&3.094&2.009\\
0.64&1.272&1.116&1.88&3.130&2.047\\
0.66&1.268&1.100&1.90&3.165&2.086\\
0.68&1.263&1.085&1.92&3.200&2.125\\
0.70&1.258&1.072&1.94&3.233&2.163\\
0.72&1.254&1.059&1.96&3.265&2.202\\
0.74&1.249&1.047&1.98&3.296&2.240\\
0.76&1.244&1.036&2.00&3.326&2.278\\
0.78&1.238&1.026&2.02&3.354&2.315\\
0.80&1.234&1.018&2.04&3.381&2.352\\
0.82&1.233&1.010&2.06&3.405&2.389\\
0.84&1.234&1.003&2.08&3.428&2.425\\
0.86&1.238&0.997&2.10&3.445&2.461\\
0.88&1.245&0.992&2.12&3.460&2.496\\
0.90&1.253&0.988&2.14&3.472&2.530\\
0.92&1.264&0.986&2.16&3.482&2.563\\
0.94&1.277&0.984&2.18&3.489&2.596\\
0.96&1.290&0.983&2.20&3.493&2.628\\
0.98&1.305&0.984&2.22&3.495&2.659\\
1.00&1.321&0.985&2.24&3.495&2.689\\
1.02&1.340&0.988&2.26&3.492&2.718\\
1.04&1.361&0.992&2.28&3.488&2.746\\
1.06&1.384&0.997&2.30&3.480&2.773\\
1.08&1.409&1.002&2.32&3.471&2.798\\
1.10&1.436&1.009&2.34&3.460&2.823\\
1.12&1.465&1.018&2.36&3.447&2.846\\
1.14&1.496&1.027&2.38&3.432&2.867\\
1.16&1.529&1.037&2.40&3.415&2.888\\
1.18&1.564&1.049&2.42&3.396&2.906\\
1.20&1.601&1.062&2.44&3.375&2.924\\
1.22&1.640&1.076&2.46&3.353&2.939\\
1.24&1.681&1.091&2.48&3.329&2.953\\
1.26&1.723&1.107&2.50&3.304&2.966\\
1.28&1.767&1.124&2.52&3.277&2.976\\
1.30&1.812&1.143&2.54&3.249&2.985\\
1.32&1.858&1.163&2.56&3.219&2.992\\
1.34&1.906&1.184&2.58&3.188&2.997\\
1.36&1.955&1.206&2.60&3.156&3.000\\
1.38&2.004&1.227&2.62&3.123&-\\
1.40&2.055&1.248&2.64&3.089&-\\
1.42&2.105&1.272&2.66&3.053&-\\
1.44&2.157&1.296&2.68&3.017&-\\
1.46&2.209&1.322&2.70&2.980&-\\
1.48&2.262&1.348&2.72&2.942&-\\
1.50&2.314&1.376&2.74&2.903&-\\
1.52&2.366&1.405&2.76&2.864&-\\
1.54&2.416&1.435&2.78&2.823&-\\
1.56&2.465&1.466&2.80&2.783&-\\
1.58&2.513&1.498&2.82&2.742&-\\
1.60&2.560&1.531&2.84&2.700&-\\
1.62&2.607&1.565&2.86&2.658&-\\[3pt]

\hline
\end{longtable}
\pagebreak

\LTcapwidth=12cm
\begin{longtable}{p{0.06\textwidth}p{0.06\textwidth}p{0.06\textwidth}p{0.06\textwidth}p{0.06\textwidth}|p{0.06\textwidth}p{0.06\textwidth}p{0.06\textwidth}p{0.06\textwidth}p{0.06\textwidth}}
\caption*{{\small{\bf Table A3}. ($P$--$X$, $Y$--$V$) and ($X$--$Y$, $Y$--$V$) relations for luminosity class V.}}\\[4pt]
\hline
\multicolumn{1}{p{0.06\textwidth}}{\newline$Y$--$V$\newline}&\multicolumn{1}{p{0.06\textwidth}}{\newline\llap{(}$P$--$X$)\rlap{$_{\rm
n}$}}&\multicolumn{1}{p{0.06\textwidth}}{\newline\llap{(}$P$--$X$)\rlap{$_{\rm
m}$}}&\multicolumn{1}{p{0.06\textwidth}}{\newline\llap{(}$X$--$Y$)\rlap{$_{\rm
n}$}}&\multicolumn{1}{p{0.06\textwidth}|}{\newline\llap{(}$X$--$Y$)\rlap{$_{\rm
m}$}}& \multicolumn{1}{p
{0.06\textwidth}}{\newline$Y$--$V$}&\multicolumn{1}{p{0.06\textwidth}}{\newline\llap{(}$P$--$X$)\rlap{$_{\rm n}$}}&\multicolumn{1}{p{0.06\textwidth}}{\newline\llap{(}$P$--$X$)\rlap{$_{\rm m}$}}&\multicolumn{1}{p{0.06\textwidth}}{\newline\llap{(}$X$--$Y$)\
rlap{$_{\rm n}$}}&\multicolumn{1}{p{0.06\textwidth}}{\newline\llap{(}$X$--$Y$)\rlap{$_{\rm m}$}} \\
[2pt] \hline
\endhead
\newline0.30&\newline0.767&\newline0.737&\newline0.490&\newline0.422&\newline0.78&\newline1.112&\newline0.795&\newline1.520&\newline1.226\\
0.32&0.744&0.700&0.502&0.423&0.80&1.126&0.830&1.573&1.269\\
0.34&0.716&0.663&0.516&0.423&0.82&1.135&0.862&1.618&1.310\\
0.36&0.686&0.625&0.532&0.422&0.84&1.140&0.890&1.654&1.350\\
0.38&0.657&0.585&0.550&0.422&0.86&1.140&0.918&1.682&1.388\\
0.40&0.631&0.544&0.570&0.423&0.88&1.137&0.943&1.705&1.423\\
0.42&0.608&0.504&0.593&0.426&0.90&1.132&0.966&1.722&1.454\\
0.44&0.592&0.464&0.618&0.433&0.92&1.125&0.985&1.733&1.482\\
0.46&0.582&0.426&0.644&0.444&0.94&1.115&1.002&1.741&1.501\\
0.48&0.580&0.391&0.675&0.459&0.96&1.105&1.014&1.744&1.513\\
0.50&0.586&0.362&0.709&0.483&0.98&1.094&1.023&1.743&1.517\\
0.52&0.600&0.341&0.748&0.516&1.00&1.083&1.027&1.739&1.510\\
0.54&0.622&0.330&0.792&0.555&1.02&1.071&1.027&1.732&1.491\\
0.56&0.652&0.333&0.839&0.599&1.04&1.059&-&1.722&-\\
0.58&0.690&0.352&0.888&0.648&1.06&1.047&-&1.711&-\\
0.60&0.733&0.393&0.941&0.703&1.08&1.034&-&1.698&-\\
0.62&0.780&0.442&0.997&0.765&1.10&1.022&-&1.685&-\\
0.64&0.831&0.495&1.058&0.831&1.12&1.009&-&1.670&-\\
0.66&0.883&0.547&1.122&0.898&1.14&0.995&-&1.656&-\\
0.68&0.934&0.595&1.190&0.959&1.16&0.981&-&1.642&-\\
0.70&0.981&0.641&1.262&1.019&1.18&0.965&-&1.628&-\\
0.72&1.024&0.683&1.334&1.075&1.20&0.948&-&1.616&-\\
0.74&1.062&0.723&1.399&1.129&1.22&0.928&-&1.606&-\\
0.76&1.090&0.760&1.461&1.179&1.24&0.906&-&1.598&-\\[3pt]
\hline
\end{longtable}

\begin{longtable}{p{0.06\textwidth}p{0.06\textwidth}p{0.06\textwidth}p{0.06\textwidth}p{0.06\textwidth}|p{0.06\textwidth}p{0.06\textwidth}p{0.06\textwidth}p{0.06\textwidth}p{0.06\textwidth}}
\caption*{{\small{\bf Table A4}. ($P$--$X$, $Y$--$V$) and ($X$--$Y$, $Y$--$V$) relations for luminosity class III.}}\\[4pt]
\hline
\multicolumn{1}{p{0.06\textwidth}}{\newline$Y$--$V$\newline}&\multicolumn{1}{p{0.06\textwidth}}{\newline\llap{(}$P$--$X$)\rlap{$_{\rm
n}$}}&\multicolumn{1}{p{0.06\textwidth}}{\newline\llap{(}$P$--$X$)\rlap{$_{\rm
m}$}}&\multicolumn{1}{p{0.06\textwidth}}{\newline\llap{(}$X$--$Y$)\rlap{$_{\rm
n}$}}&\multicolumn{1}{p{0.06\textwidth}|}{\newline\llap{(}$X$--$Y$)\rlap{$_{\rm
m}$}}& \multicolumn{1}
{p{0.06\textwidth}}{\newline$Y$--$V$}&\multicolumn{1}{p{0.06\textwidth}}{\newline\llap{(}$P$--$X$)\rlap{$_{\rm n}$}}&\multicolumn{1}{p{0.06\textwidth}}{\newline\llap{(}$P$--$X$)\rlap{$_{\rm m}$}}&\multicolumn{1}{p{0.06\textwidth}}{\newline\llap{(}$X$--$Y$
)\rlap{$_{\rm n}$}}&\multicolumn{1}{p{0.06\textwidth}}{\newline\llap{(}$X$--$Y$)\rlap{$_{\rm m}$}} \\
[2pt] \hline
\endhead
\newline0.20&\newline0.847&\newline-&\newline0.347&\newline-&\newline0.78&\newline1.017&\newline0.566&\newline1.330&\newline0.834\\
0.22&0.847&-&0.378&-&0.80&1.058&0.585&1.383&0.862\\
0.24&0.840&-&0.407&-&0.82&1.098&0.606&1.435&0.889\\
0.26&0.827&-&0.433&-&0.84&1.137&0.631&1.489&0.917\\
0.28&0.810&-&0.457&-&0.86&1.173&0.656&1.542&0.945\\
0.30&0.789&0.730&0.480&0.388&0.88&1.207&0.682&1.596&0.972\\
0.32&0.766&0.691&0.501&0.392&0.90&1.239&0.709&1.649&0.998\\
0.34&0.741&0.656&0.522&0.398&0.92&1.268&0.736&1.702&1.024\\
0.36&0.715&0.624&0.542&0.406&0.94&1.295&0.763&1.754&1.050\\
0.38&0.690&0.595&0.562&0.415&0.96&1.318&0.790&1.805&1.075\\
0.40&0.667&0.569&0.582&0.426&0.98&1.338&0.817&1.854&1.101\\
0.42&0.645&0.546&0.602&0.438&1.00&1.356&0.843&1.901&1.127\\
0.44&0.627&0.527&0.624&0.452&1.02&1.371&0.870&1.946&1.153\\
0.46&0.613&0.510&0.650&0.467&1.04&1.383&0.895&1.987&1.179\\
0.48&0.603&0.496&0.678&0.483&1.06&1.393&0.920&2.023&1.204\\
0.50&0.599&0.485&0.708&0.500&1.08&1.400&0.945&2.049&1.230\\
0.52&0.601&0.476&0.741&0.519&1.10&1.405&0.968&2.068&1.256\\
0.54&0.608&0.470&0.775&0.539&1.12&1.409&0.990&2.081&1.282\\
0.56&0.622&0.466&0.812&0.559&1.14&1.411&1.011&2.086&1.308\\
0.58&0.641&0.465&0.852&0.581&1.16&1.412&1.030&2.084&1.333\\
0.60&0.665&0.466&0.893&0.604&1.18&1.409&1.048&2.075&1.359\\
0.62&0.694&0.469&0.937&0.627&1.20&1.404&1.064&2.059&1.385\\
0.64&0.728&0.475&0.982&0.651&1.22&1.399&1.079&2.035&1.411\\
0.66&0.766&0.482&1.030&0.676&1.24&1.395&1.091&2.003&1.436\\
0.68&0.806&0.491&1.078&0.701&1.26&1.392&1.101&1.963&1.462\\
0.70&0.848&0.503&1.128&0.727&1.28&1.389&1.109&1.915&1.488\\
0.72&0.890&0.516&1.177&0.753&1.30&1.386&1.114&1.858&1.514\\
0.74&0.933&0.531&1.228&0.780&1.32&1.385&-&1.792&-\\
0.76&0.975&0.547&1.279&0.807&1.34&1.384&-&1.717&-\\[3pt]
\hline
\end{longtable}

\end{document}